\documentclass[12pt]{iopart}
\usepackage{graphicx}
\usepackage{amssymb}
\usepackage{color}
\input colors.defs

\newcommand{\bminil}[1]{\begin{minipage}[l]{#1 \textwidth}}
\newcommand{\bminir}[1]{\begin{minipage}[r]{#1 \textwidth}}
\newcommand{\bminic}[1]{\begin{minipage}[c]{#1 \textwidth}}
\newcommand{\emini}{\end{minipage}}
\newcommand{\EQL}{\begin{equation}\label}
\newcommand{\EQ}{\begin{equation}}
\newcommand{\EN}{\end{equation}}
\newcommand{\ITM}{\begin{itemize}}
\newcommand{\ITN}{\end{itemize}}
\newcommand{\ENM}{\begin{enumerate}}
\newcommand{\EEN}{\end{enumerate}}
\def\DEL#1{\textcolor{Green}{}}     
\def\ADD#1{}         
\def\AD#1{{\textcolor{magenta}{}}}       
\def\RoraiQ#1{{\textcolor{cyan}{}}}         
\def\RMKA#1{}         
\newcommand{\dfrac}[2]{{\displaystyle\frac{#1}{t}}}

\newcommand{\btriangle}{\mbox{\boldmath$\triangle$}}

\newcommand{\bB}{\mbox{\boldmath$B$}}

\newcommand{\bN}{\mbox{\boldmath$N$}}

\newcommand{\bS}{\mbox{\boldmath$S$}} 
\newcommand{\bT}{\mbox{\boldmath$T$}}

\newcommand{\bk}{\mbox{\boldmath$k$}}

\newcommand{\bbr}{\mbox{\boldmath$r$}}

\newcommand{\bu}{\mbox{\boldmath$u$}}

\newcommand{\bx}{\mbox{\boldmath$x$}}

\newcommand{\bbR}{\mathbb{R}}

\newcommand{\bomega}{\mbox{\boldmath$\omega$}}

\newcommand{\p}{\partial}

\newcommand{\ppto}[1]{\frac{\partial #1}{\partial t}}

\newcommand{\half}{\mbox{$\frac{1}{2}$}}

\newcommand{\sbl}[3]{\|{#1}\|_{#2}^{#3}} 
\newcommand{\eqref}[1]{(\ref{#1})} 
\newcommand{\etall}{{\it et al.}}
\newcommand{\biband}{and~}
\newcommand{\authone}[2]{#2~#1}
\newcommand{\authtwo}[4]{#2~#1~\biband~#4~#3}
\newcommand{\auththr}[6]{#2~#1,~#4~#3~\biband~#6~#5}
\newcommand{\authfour}[8]{#2~#1,~#4~#3~#6~#5~\biband~#8~#7} 
\newcommand{\authmanytwo}[4]{#2~#1~#4~#3}

\newcommand{\yearp}[2]{(#1 #2)}
\newcommand{\yearpp}[1]{(#1)}
\newcommand{\KlecknerIrvinethirt}{Kleckner \biband Irvine~} 
\newcommand{\ScheeleretalIrvinefourt}{Scheeler \etall~} 
\newcommand{\BiferaleKerrninefive}{Biferale \biband Kerr~} 
\newcommand{\Calugareanufivenine}{Calugareanu~} 
\newcommand{\Constantineightsix}{Constantin~} 
\newcommand{\Laingetalfivet}{Laing \etall~} 
\newcommand{\HoughtonCarpenterthr}{Houghton \biband Carpenter~} 
\newcommand{\RMK}{Kerr~} 
\newcommand{\Moffattfourt}{Moffatt~} 
\newcommand{\MoffattRiccaninetwo}{Moffatt \biband Ricca~} 
\newcommand{\Sahooetalfivet}{Sahoo \etall~} 

\newcommand{\yjour}[6]{~#1~\textit{#2} \textbf{#3}, #4#5.}

\newcommand{\ybook}[3]{~#1~ \textit{#2}. #3.}

\begin{document}

\title{Trefoil knot structure during reconnection}
%
\author{Robert M. Kerr}
\address{Department of Mathematics, University of Warwick, United Kingdom}

\ead{Robert.Kerr@warwick.ac.uk\\ 22 May 2017}
\begin{abstract}
Three-dimensional images of evolving numerical trefoil vortex knots are used to study the 
growth and decay of the enstrophy and helicity.  Negative helicity density ($h<0$) plays 
several roles.  First, during anti-parallel reconnection, sheets of oppositely-signed helicity 
dissipation of equal magnitude on either side of the maximum of the enstrophy dissipation allow 
the global helicity ${\cal H}$ to be preserved through the first reconnection, as 
suggested theoretically \yearp{\Laingetalfivet}{2015} and observed experimentally 
\yearp{\ScheeleretalIrvinefourt}{2014}. Next, to maintain the growth of the enstrophy and positive
helicity within the trefoil while ${\cal H}$ is preserved, $h<0$ forms in the outer parts
of the trefoil so long as the periodic boundaries do not interfere. To prevent that,
the domain size $\ell$ is increased as the viscosity $\nu\to0$.
Combined, this allows two sets of trefoils to form a new scaling
regime with linearly decreasing $(\sqrt{\nu}Z(t))^{-1/2}$ up to common $\nu$-independent
times $t_x$ that the graphics show is when the first reconnection ends.
During this phase there is good correspondence between the 
evolution of the simulated vortices and the reconnecting experimental trefoil of 
\KlecknerIrvinethirt\yearpp{2013} when time is scaled by their respective nonlinear timescales 
$t_f$. The timescales $t_f$ are based upon by the radii $r_f$ of the trefoils and their circulations 
$\Gamma$, so long as the strong camber of the experimental hydrofoil models is used to correct
the published experimental circulations $\Gamma$ that use only the flat-plate approximation.
\end{abstract}
%
%

%
%
Even though most turbulent flows are neither homogeneous, isotropic or statistically steady,
most of our undertanding of turbulence uses those assumptions. Part of the reason is that 
for both experiments and simulations it is difficult to inject unstable, inherently anisotropic 
configurations into flows. In particular, flows that are free of the effects of walls or periodicity. 
\KlecknerIrvinethirt\yearpp{2013} have shown how this can now be done experimentally by 
3D-printing hydrofoil knots, either linked rings or trefoils, covering them with hydrogen bubbles,
then yanking them out of a water tank. This creates helical vortex knots whose low pressure vortex 
cores are marked by strings of bubbles.

\ScheeleretalIrvinefourt\yearpp{2014} have extended this repertoire by showing how 
the trajectories of bubbles can be used to determine the evolution of their centreline helicities,
a measure of topological helicity based upon how those trajectories cross one another
from several perspectives. The advantage of determining the topology using this diagnostic 
is that it does not require
direct measurements of the velocity $\bu$ and vorticity $\bomega$ used by the continuum
helicity density $h=\bu\cdot\bomega$ \eqref{eq:helicity}. How the topology of bubble trajectories 
changes also provides a diagnostic for setting the timescales. 

The surprising result for the evolution of the centreline helicity for the trefoil 
in \ScheeleretalIrvinefourt\yearpp{2014} is that this diagnostic
is preserved despite clear signs that the topology of the trefoil is changing due to vortex 
reconnection. Would the true continuum helicity be preserved in numerical simulations that 
qualitatively reproduce the observed topological changes of the evolving experimental trefoils? 

This paper will use two sets of high-resolution, strongly perturbed trefoils from\RMK\yearpp{2017}  
to identify how physical space structures reconnect during a period for which the continuum global
helicity ${\cal H}$ \eqref{eq:helicity} is similarly preserved.  \RMK\yearpp{2017} also 
identified a new scaling regime for the growth of $\sqrt{\nu}Z(t)$, $\nu$ the 
viscosity and $Z$ the volume-integrated enstrophy \eqref{eq:enstrophy}, that
was first indicated by $\nu$-independent crossing times $t_x$ with a common $\sqrt{\nu}Z(t_x)$. 
What is the role of the structures in the dynamics responsible for this unexpected behaviour of
both the helicity ${\cal H}$ and enstrophy $Z$ over this period?
The plan is to use three-dimensional graphics from a time just before reconnection begins 
to the time, after reconnection ends, when helicity finally begins to decay to address that 
question.

The trefoils' initial trajectories are illustrated in figure \ref{fig:T6} and figure \ref{fig:QSsnuZ} 
summarises the $\sqrt{\nu}Z$ scaling using 
\EQL{eq:Bnu} B_\nu(t)=\Bigl(\sqrt{\nu}Z(t)\Bigr)^{-1/2}\quad{\rm with}~\nu{\rm-independent}\quad 
B_x=B_\nu(t_x)\quad{\rm at}~t_x\,.\EN
The characteristic crossing times are $t_x$(Q)=40 and $t_x$(S)$\approx$45 for the Q and S-trefoil 
calculations listed in table \ref{tab:cases} and $B_\nu(t)$ is linearly decreasing in both subplots
for $t_\Gamma\!\leq \!t\!\leq\! t_x$, $t_\Gamma\approx15$. 
By linearly extrapolated this behaviour to critical times $T_c(\nu)> t_x$ using 
\EQL{eq:TcDt} T_c(\nu)= \frac{t_x-t_\Gamma B_{x/\Gamma}(\nu)}{1-B_{x/\Gamma}(\nu)} 
\quad{\rm where}\quad B_{x/\Gamma}(\nu)=B_x/B_\nu(t_\Gamma)\,, \EN
self-similar collapse was identified over the times $t_\Gamma\leq t\leq t_x$ \yearp{\RMK}{2017} using
\EQL{eq:isnuZBxtime} (T_c(\nu)-t_x)\bigl(B_\nu(t)-B_x\bigr)=
(T_c(\nu)-t_x)\left(\Bigl(\sqrt{\nu}Z(t)\Bigr)^{-1/2}-B_x \right) \,.\EN

\RMK\yearpp{2017} also showed that this self-similar collapse could be applied to new anti-parallel 
calculations and provided three-dimensional graphics to show that the anti-parallel $t_\Gamma$ is
when reconnection began by exchanging circulation $\Gamma$ between the vortices and that the
$t_x$ determined by the crossings of $B_\nu(t)$ is when those reconnections finish. 
It will be shown in section \ref{sec:t40} that $t_x$(Q)=40 also 
represents the end of the first Q-trefoil reconnection and is the best timescale for comparisons 
with the experiments  of \KlecknerIrvinethirt\yearpp{2013}.

The comparisons between the Q-trefoil images and those from \KlecknerIrvinethirt\yearpp{2013} 
are feasible because the perturbed trefoils, illustrated in figure \ref{fig:T6}, 
were constructed so that there would be a single dominant initial reconnection, 
as in the experiments, with the comparisons around the reconnection time $t\sim t_x$ 
discussed in section \ref{sec:t40}. Establishing those similarities justifies using
the simulations to provide diagnostics that are inaccessible to those experiments, but needed
for explaining the observed dynamics.
For example, figure \ref{fig:T36} shows the terms  in helicity budget equation \eqref{eq:helicity} 
that might allow preservation of the global helicity.

How do the global helicity ${\cal H}(t)$ and energy dissipation rate $\epsilon(t)=\nu Z$ 
\eqref{eq:energy} evolve after $t=t_x$?  For both trefoils and before $t\sim 2t_x$, ${\cal H}(t)$ 
begins to decay in figure \ref{fig:HelQS} and for the dissipation rate,
figure \ref{fig:QSsnuZ} indicates fixed times $t_\epsilon$ (stars) 
when the $\nu$-independent dissipation rates $\epsilon(t)$ start to saturate for $t>t_\epsilon$.  
The constant $\epsilon(t_\epsilon)=\nu Z(t_\epsilon)$ and 
earlier constant $\sqrt{\nu}Z(t_x)$ 
both imply that $Z\!\to\!\infty$ as $\nu\!\to\!0$.  Can this growth in the enstrophy $Z$ 
be maintained as $\nu\!\to\!0$, eventually leading to a {\it dissipation anomaly}?
That is, can there be finite energy dissipation in a finite time as $\nu\!\to\!0$? 

To answer that, \RMK\yearpp{2017} considered the effect of accepted Sobolev space mathematical 
analysis for smooth solutions of the Navier-Stokes equation that shows that if the domain size 
$V=\ell^3$ is fixed, then $Z(t)$ has an upper bound as $\nu\!\to\!0$ 
\yearp{\Constantineightsix}{1986}.
However, this mathematics also allows this constraint to be relaxed by increasing $\ell$. 
This escape valve, increasing the domain size $V=\ell^3$ as $\nu\!\to\!0$, 
was used for both the trefoils 
and the anti-parallel vortices to ensure that the collapse defined by \eqref{eq:isnuZBxtime}
could be maintained as the viscosity $\nu$ was decreased. 
Is there physical space dynamics that 
underlies this rigorously derived constraint and its relaxation? The late time graphics in 
section \ref{sec:latetimes} will suggest that these effects could arise from the generation of 
negative helicity in the outer parts of the trefoil domain.

The thinner-core S-trefoils have been included to provide continuum helicity ${\cal H}$ comparisons 
to the centreline helicity diagnostics of the thinner vortex core trefoil experiment of 
\ScheeleretalIrvinefourt\yearpp{2014}, which \KlecknerIrvinethirt\yearpp{2013} did not provide. 
The connection to the physical space diagnostics from \KlecknerIrvinethirt\yearpp{2013} will be 
provided by the similarities between the Q and S-trefoils for the evolution of $B_\nu(t)$ 
\eqref{eq:Bnu} and global helicity \eqref{eq:helicity} in figures \ref{fig:QSsnuZ} 
and \ref{fig:HelQS}.

This paper is organised as follows. After introducing the equations, diagnostics and
initialisation, illustrated for the Q-trefoil in figure \ref{fig:T6}, 
there is a short review of the $\sqrt{\nu}Z$ enstrophy scaling results in \RMK\yearpp{2017}. 
Then a re-evaluation of the circulations reported for the experiments and the new timescales these 
imply.  Once these are established, then graphics and analysis connecting the dynamics of positive 
helicity and enstrophy production and the appearance of negative helicity increasingly far from the 
original trefoil will be given. At the end, structural changes during the last period when helicity 
finally begins to decay and finite energy dissipation is generated are presented.

\section{Equations, diagnostics and initial condition\label{sec:diagnostics}}

The governing equations in this paper will be the incompressible 
Navier-Stokes velocity equations 
\EQL{eq:NS} \hspace{-8mm} \ppto{\bu} + ({\bu}\cdot\nabla){\bu} = -\nabla p+ 
\underbrace{\nu\triangle{\bu}}_{\rm dissipation}, \qquad 
\nabla\cdot{\bu}=0\EN
and the diagnostics will primarily use the vorticity $\bomega=\nabla\times\bu$ , which obeys
\EQL{eq:omega} \hspace{-8mm} \ppto{\bomega} + ({\bu}\cdot\nabla){\bomega} = 
({\bomega}\cdot\nabla){\bu} + \nu\triangle{\bomega},\qquad \nabla\cdot{\bomega}=0\,.\EN
All of the calculations were done in periodic boxes of variable size ($V=\ell^3$). 
The continuum equations for the densities of the energy, enstrophy and helicity,
$e=\half|\bu|^2$, $|\bomega|^2$ and $h=\bu\cdot\bomega$ respectively are (with their 
volume-integrated measures):
\EQL{eq:energy} \hspace{-15mm} \ppto{e}+ ({\bu}\cdot\nabla)e = -\nabla\cdot(\bu p) 
+\nu\triangle e
-\underbrace{\nu(\nabla\bu)^2}_{\epsilon={\rm dissipation}=\nu Z},\qquad 
E=\half\int\bu^2dV\,.\EN
\EQL{eq:enstrophy} \hspace{-15mm} \ppto{|\bomega|^2}+ ({\bu}\cdot\nabla)|\bomega|^2 = 
\underbrace{2\bomega\bS\bomega}_{Z_p={\rm production}}
+\nu\triangle|\bomega|^2
-  \underbrace{2\nu(\nabla\bomega)^2}_{\epsilon_\omega=Z-{\rm dissipation}},\qquad 
Z=\int\bomega^2dV\,.\EN
\EQL{eq:helicity} \hspace{-15mm} \ppto{h}+ ({\bu}\cdot\nabla)h = 
\underbrace{-\bomega\cdot\nabla\Pi}_{\omega-{\rm transport}}
+\underbrace{\nu\btriangle h}_{\nu-{\rm transport}} -\underbrace{
2\nu{\rm tr}(\nabla\bomega\cdot\nabla\bu^T)}_{\epsilon_h={\cal H}-{\rm dissipation}}
\qquad
{\cal H}=\int\bu\cdot\bomega dV\,.\EN
$\Pi=p-\half\bu^2\neq p_h$ is not the pressure head $p_h=p+\half\bu^2$.
Both the global energy $E$ and helicity ${\cal H}$ are inviscid invariants, but the
local helicity density $h$ is not locally Galilean invariant due to the $\omega$-transport term
and its role in nonlinearity is not fully understood \yearp{\Moffattfourt}{2014}. The global
helicity ${\cal H}$ has same dimensional units as the circulation-squared.  

The circulations $\Gamma_i$ about vortices with distinct trajectories $\bx_i \in{\cal C}_i$ are
another set of inviscid invariants:
\EQL{eq:Gamma} \Gamma_i=\oint \bu_i\cdot d\bbr_i \quad{\rm where}\quad \bbr_i~~
\mbox{is a closed loop about}~~{\cal C}_i\,. \EN
Under Navier-Stokes dynamics, these circulations will change as vortices of opposite sign
meet and begin to reconnect in a gradual process where, through the viscous terms,
new vortices are generated as the original vortices annihilate one another, as demonstrated
by figure \ref{fig:T31}.

Under Navier-Stokes dynamics, the local helicity density can be of either sign, 
can grow, decrease and even change sign due to both the viscous terms and the $\omega$-transport 
term along the vortices in \eqref{eq:helicity}. Also, unlike the kinetic energy which
cascades overwhelmingly to small scales, $h$ can move to both large and 
small scales (\BiferaleKerrninefive 1995, \Sahooetalfivet 2015). 

The diagnostics will be continuum properties in the equations above 
(\ref{eq:energy},\ref{eq:enstrophy},\ref{eq:helicity}) and a few discrete vortex trajectories
\eqref{eq:vortexlines}. Of these,
the most important vorticity diagnostic will be the enstrophy $Z$, which can grow inviscidly due 
to its production term $Z_p$. The maximum of vorticity magnitude will be denoted
\EQL{eq:omegam} \sbl{\omega}{\infty}{}=\sup|\bomega| \EN
and its location $\bx_\infty$ will be used to seed the trajectories of vortex lines. 
However, $\|\omega\|_\infty$  will not have its usual H\"older significance as a volume-averaged 
upper bound for all norms because $Z$ and ${\cal H}$ will be volume-integrated norms,

\subsection{Vortex lines and linking numbers\label{sec:link}}

To provide qualitative comparisons with the experimental vortex lines 
(\KlecknerIrvinethirt 2013, \ScheeleretalIrvinefourt 2014), 
vortex lines $\bx_j(s)$ were identified by solving the 
following ordinary differential equation using the Matlab streamline function, 
\EQL{eq:vortexlines} \frac{d\bx_j(s)}{ds}=\bomega(\bx_j(s))\,. \EN
The seeds for solving \eqref{eq:vortexlines} were chosen from the positions around, but not 
necessarily at, local vorticity maxima. 

When the vortices are distinct with closed trajectories, 
these vortex lines have the following topological 
numbers: The integer linking numbers ${\cal L}_{ij}$ between all distinct vortex trajectories, 
the integer self-linking numbers ${\cal L}_{Si}=W\!r_i+T\!w_i$ of individual closed loops, and 
the non-integer writhe and twist $W\!r_i$ and $T\!w_i$. Then, by assigning
circulations $\Gamma_i$ to the vortices and summing one can determine the global helicity
${\cal H}$ \yearp{\MoffattRiccaninetwo}{1992}. 
\EQL{eq:Hlink} {\cal H}=\sum_{ij}\Gamma_i\Gamma_j{\cal L}_{ij}
+\sum_{i}\Gamma_i^2 {\cal L}_{S_i}\,.\EN

\begin{figure} 
\includegraphics[scale=0.32]{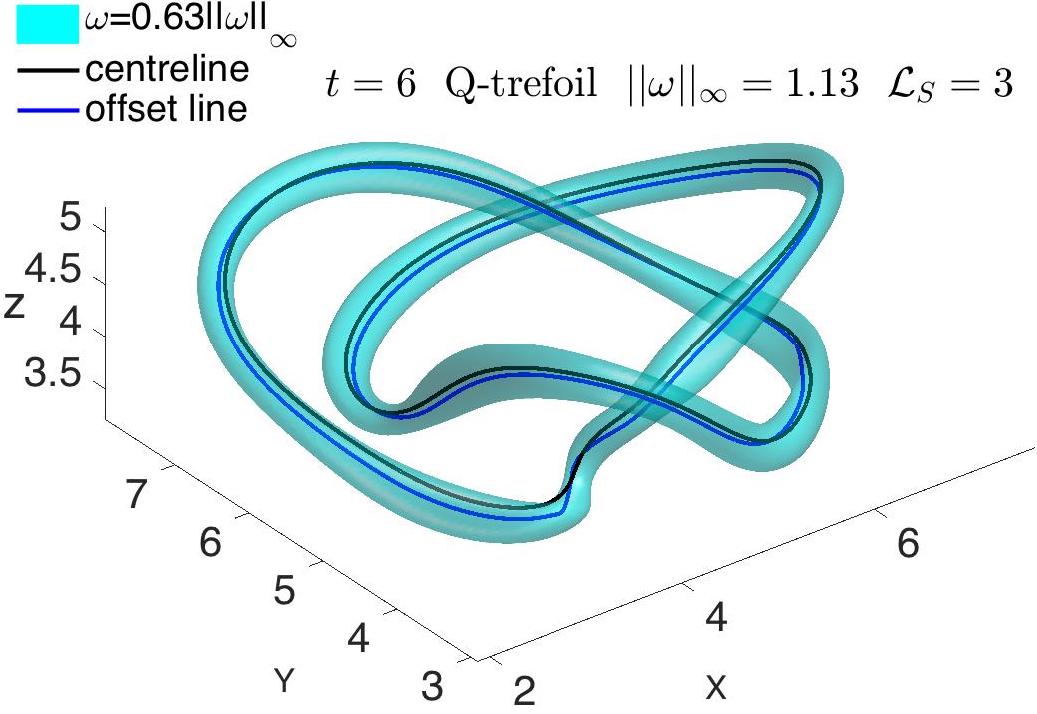}
\begin{picture}(0,0)\put(-60,32){\huge $t=6$}\end{picture}
\caption{\label{fig:T6} Vorticity isosurface plus two closed vortex lines
of the perturbed trefoil vortex at $t=6$, not long after initialization. 
Its self-linking is ${\cal L}_S=3$, which can be split into writhe of $W\!r=3.15$ 
and twist of $T\!w=-0.15$.}
\end{figure}
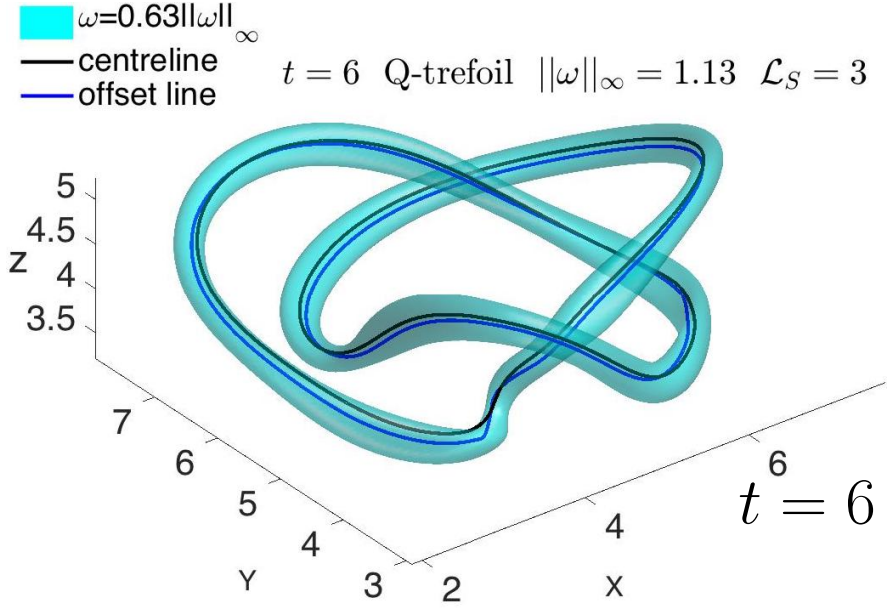

The quantitative tool that is used to determine the writhe, self-linking and
intervortex linking numbers in figure \ref{fig:T31} is a regularised Gauss linking integral 
about two loops $\bx_{i}\in{\cal C}_{i}$ and $\bx_{j}\in{\cal C}_{j}$
\EQL{eq:link} {\cal L}_{ij} =
\sum_{ij}\frac{1}{4\pi}\oint_{{\cal C}_i}\oint_{{\cal C}_j}
\frac{(d\bx_i\times d\bx_j)\cdot(\bx_i-\bx_j)}{(|\bx_i-\bx_j|^2+\delta^2)^{1.5}}\,. 
\EN
The regularisation of the denominator using $\delta$ has been added for determining the
writhe when $i=j$ (\Calugareanufivenine 1959,\MoffattRiccaninetwo 1992), with $\delta\equiv 0$
for directly determining the self-linking numbers ${\cal L}_{Si}$ using 
two parallel trajectories within the vortex cores, as illustrated in figure \ref{fig:T6}, 
and for the intervortex linking numbers with $i\neq j$

Defining the Frenet-Serret relations of a curve in space: $\bx(s):[0,1]\rightarrow\bbR^3$ as
\EQL{eq:FS} 
\bT(s) =\p_s\bx_j(s);\quad\p_s\bN =\tau\bB-\kappa\bT;\quad
\kappa \bN  =\p_s\bT;\quad\p_s\bB =-\tau\bN \,,
\EN
where the $\kappa$ is the curvature and $\tau$ is the torsion,
the intrinsic twist of a closed loop $T\!w_i$ can in principle be
determined from the line integral of the torsion of the vortex lines:
\EQL{eq:twist} T\!w_i=\frac{1}{2\pi}\oint \tau ds,\qquad{\rm where}\quad
\tau=\frac{d\bN}{ds}\cdot\bB\,.\EN
Because determining $\tau$ requires taking third-derivatives of the positions $\bx(s)$, 
which the single-precision analysis data used here cannot determine accurately, the values of
the twist will always be $T\!w={\cal L}_S-W\!r$. 

\begin{table}
  \begin{center}
\begin{tabular}{ccllcccccc}
Cases & Domains & $r_0$ & $\omega_{\rm in}$ & $k_f$ & $r_e$ & $\omega_0$ & $Z_0$ 
& $E_0$ &$\nu$ \\
Q& $(3\pi)^3$   &0.25& 1.26  & 11.9 &0.40 & 1 & $5.48$  & 0.96
&5e-4 to 1.25e-4 \\
Q& $(4\pi)^3$   &0.25& 1.26  & 11.9 &0.40 & 1 & $5.29$ & 0.85 & $5\times10^{-4}$ to 
$6.25\times10^{-5}$\\
Q& $(6\pi)^3$   &0.25& 1.26  & 11.9 &0.40 & 1 & $5.30$ & 0.90 & $3.125\times10^{-5}$\\
S& $(6\pi)^3$   &0.125& 5  & 23.8 &0.20 & 4 & $17.7$ & 1.03 & (2.5 and 1.25)$\times10^{-4}$ \\
S& $(9\pi)^3$ &0.125& 5  & 23.8 &0.20 & 4 & $17.8$ & 1.13-1.17 & (6.25 and 3.125)$\times10^{-5}$ \\
S& $(12\pi)^3$ &0.125& 5  & 23.8 &0.20 & 4 & $17.8$ & 1.13-1.17 & $7.8\times10^{-6}$ \\
\end{tabular}
\caption{Parameters for the initial conditions using \eqref{eq:Rosenheadfilter}
and viscosities of the cases. 
$r_0$ and $\omega_{\rm in}$ are the pre-filter radius and vorticity of the filaments.
$k_f$ are the filter wavenumbers that give initial conditions with
$\omega_0=\|\bomega\|_\infty$, enstrophy=$Z_0$ and energy=$E_0$. 
The final meshes for the $(3\pi)^3$ calculations were $512^3$, for the
$\nu\geq1.25\times10^{-4}$ $(4\pi)^3$ calculations were $1024^3$ and for the rest $2048^3$.
The initial helicity for all of the calculations is 
${\cal H}(t=0)=7.67\times10^{-4}$ for circulations of $\Gamma=0.505$. 
The $(3\pi)^3$-Q cases are used only for graphics.
}
  \label{tab:cases}
  \end{center}
\end{table}

\subsection{Initial condition and length scales \label{sec:initial}}

The initial trajectory of all of the simulated trefoils will be
\EQL{eq:trefoil}\begin{array}{rrl} x(\phi)= r(\phi)\cos(\alpha) &\quad
y(\phi)= & r(\phi)\sin(\alpha) \qquad z(\phi)= a\cos(\alpha) \\
{\rm where} & r(\phi) =& r_f+r_1a\cos(\phi) +a\sin(w\phi+\phi_0)\\
{\rm and} &  \alpha=& \phi+a\cos(w\phi+\phi_0)/(wr_f) \end{array} \EN
with $r_f=2$, $a=0.5$, $w=1.5$, $\phi_0=0$, $r_1=0.25$ and $\phi=[1:4\pi]$.  This 
weave winds itself twice about the central deformed ring with: $x_c^2(\phi)+y_c^2(\phi)=r_c^2(\phi)$,
where $r_c(\phi)=r_f+r_1a\cos(\phi)$  for a $r_1\neq0$ perturbation. 
The separation through the $r=r_c$ ring of the two loops of the trefoil is $\delta_a=2a=1$. 
Four additional low intensity vortex rings, two moving up in $z$ and two down, 
provided the perturbation that breaks the three-fold symmetry of the trefoil 
so that it has a single major initial reconnection like the experiments.  
Once all the trajectories are defined, 
vortices of finite radii $r_0$ are mapped from this trajectory onto the mesh using a profile
function based upon the Rosenhead regularisation of a point vortex, then the fields on the mesh 
are smoothed with a hyperviscous filter, as described previously (\RMK 2013, \RMK 2017).
\EQL{eq:Rosenheadfilter} |\omega_i|(r)=\Gamma\frac{(r_0^2)}{(r^2+r_0^2)^2},\quad
\omega_f(\bk)=\omega_i(\bk)\exp\left(-\frac{k^4}{k_f^4}\right),\quad 
r_e=\left(\frac{\Gamma}{\omega_0/\pi}\right)^{1/2} \EN where the $r_e$ in table \ref{tab:cases} are the radii of the smoothed vortices.
All of the trefoils have a circulation of $\Gamma=0.505$
and after filtering their effective radii all obey $r_e\approx1.6 r_0$.

The two relevant length scales are $r_f=2$, the trefoil's radius, and $r_e$, the effective
thickness of the filaments.  Two $r_e$ are listed in table \ref{tab:cases}, 
designated Q and S, with $r_e(S)=0.5r_e(Q)$, so that the effective cross-sectional areas 
of the vortices, $A(r_e)=\pi r_e^2$, are $A(r_e(S))=\pi r_e^2(S)=0.25A(r_e(Q))=0.25\pi r_e^2(Q)$.
To keep the circulations $\Gamma$ constant,
the initial centreline vorticity $\|\omega\|_\infty(t=0)$ changes inversely with the $A(r_e)$.

\begin{figure}
\includegraphics[scale=0.21]{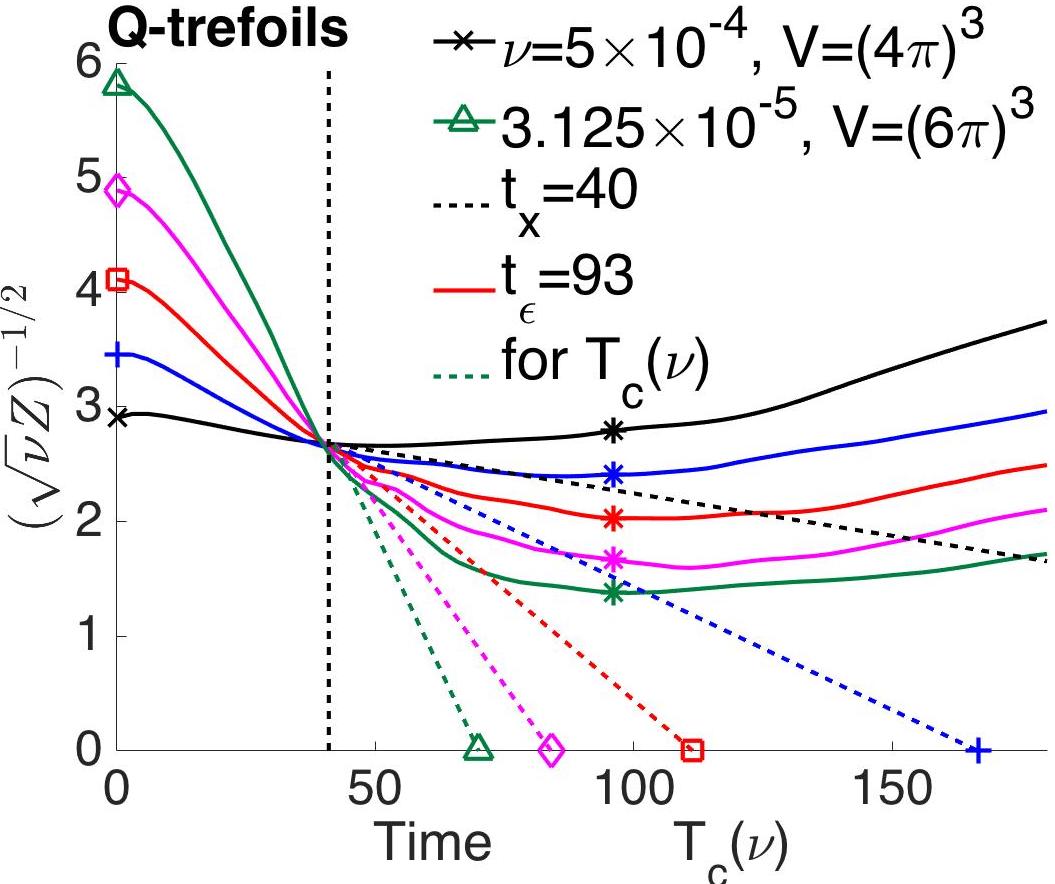}
\includegraphics[scale=0.21]{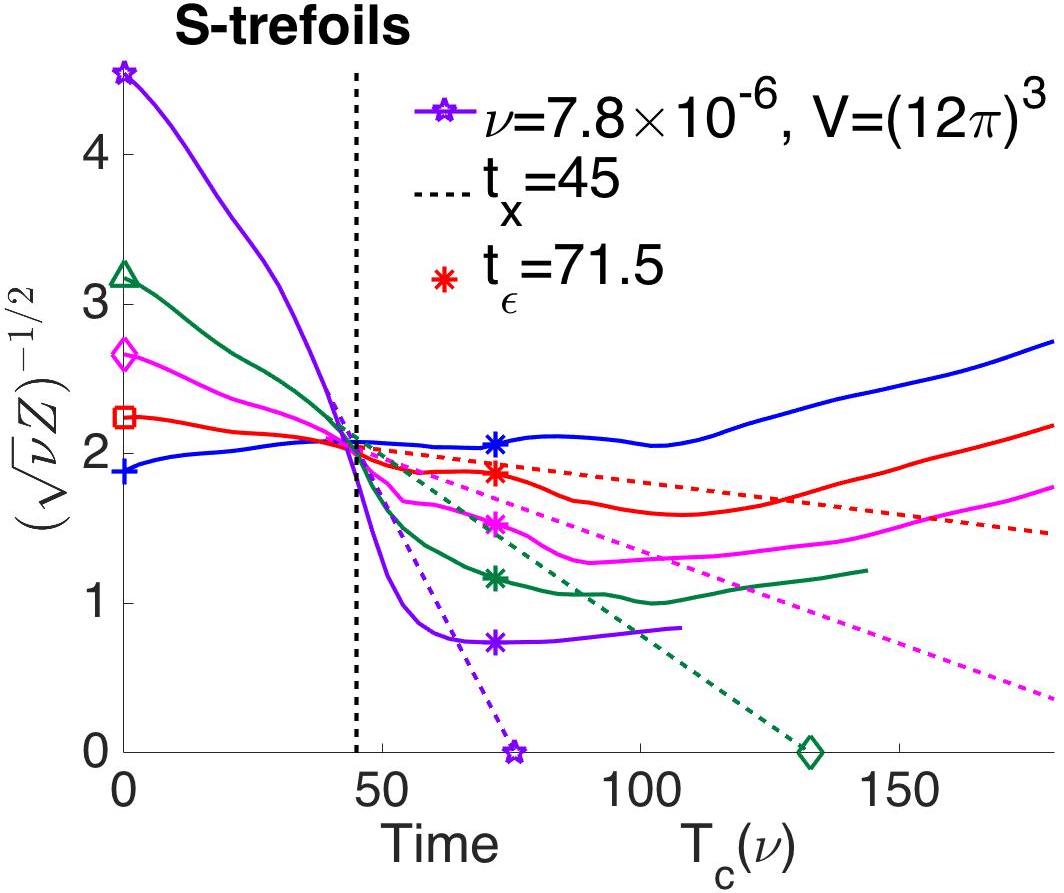}
\begin{picture}(0,0)\put(160,135){{\LARGE\bf (a)}}
\put(380,135){{\LARGE\bf (b)}}\end{picture}
\caption{\label{fig:QSsnuZ} 
Scaling of $B_\nu(t)=(\sqrt{\nu}Z(t))^{-1/2}$ \eqref{eq:Bnu}, inverse, 
$\sqrt{\nu}$-scaled enstrophies with $r_e$ dependent crossing times $t_x$ 
for the cases in table \ref{tab:cases}.  
Colours for the different viscosities are the same as in figure \ref{fig:HelQS} 
Consistent decreasing $B_\nu(t)$ begins at $t_\Gamma\approx15$ in each subplot and the
dashed lines indicate the linear extrapolations 
that determine the effective critical times $T_c(\nu)$ \eqref{eq:TcDt}.
The $*$'s indicate when the energy dissipation rates $\epsilon=\nu Z$ begin 
to saturate at $\nu$-independent values of $\epsilon$.
{\bf a:} Q-trefoils with $t_x\approx40$ for viscosities that 
vary by decreasing factors of 2 from $\nu=5\times10^4$ (black) to $\nu=$3.125e-5 (green). 
The linearly-decreasing behaviour of $B_\nu(t)$ begins at $t_\Gamma\approx15$ as 
negative helicity, as shown in figure \ref{fig:T24}, first  appears in the graphics. 
\RMK\yearpp{2017} used these $T_c(\nu)$ in \eqref{eq:isnuZBxtime} 
to find a self-similar collapse of the $B_\nu(t)$.
{\bf b:} S-trefoils with $t_x\approx45$. The viscosities are those in figure \ref{fig:HelQS}
plus $\nu=2.5\times10^{-4}$ and $6.125\times10^{-5}$. $t_\Gamma=25$ was used for the
$T_c(\nu)$ linear extrapolations.
}
\end{figure}
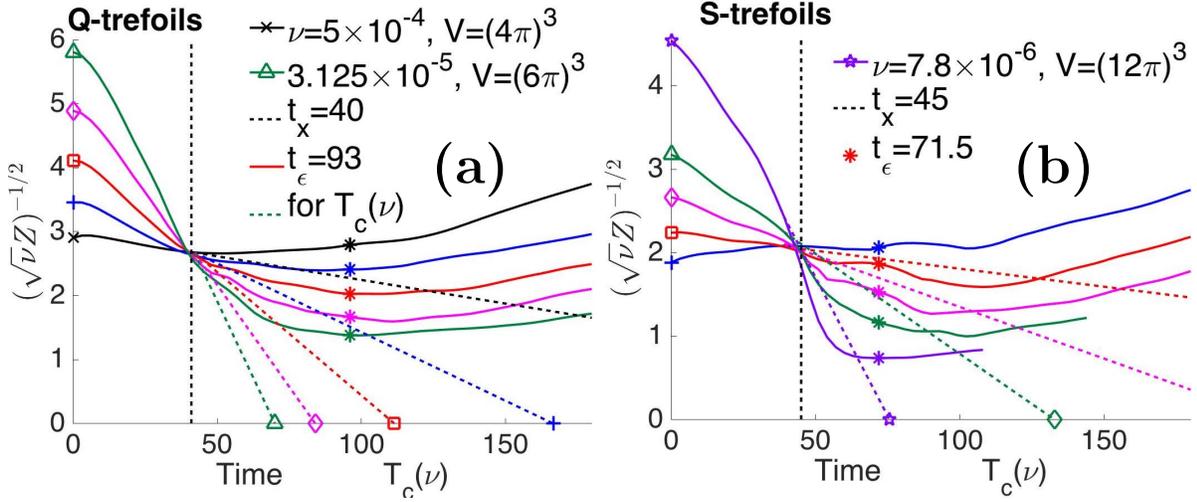

\begin{figure} 
\includegraphics[scale=0.36]{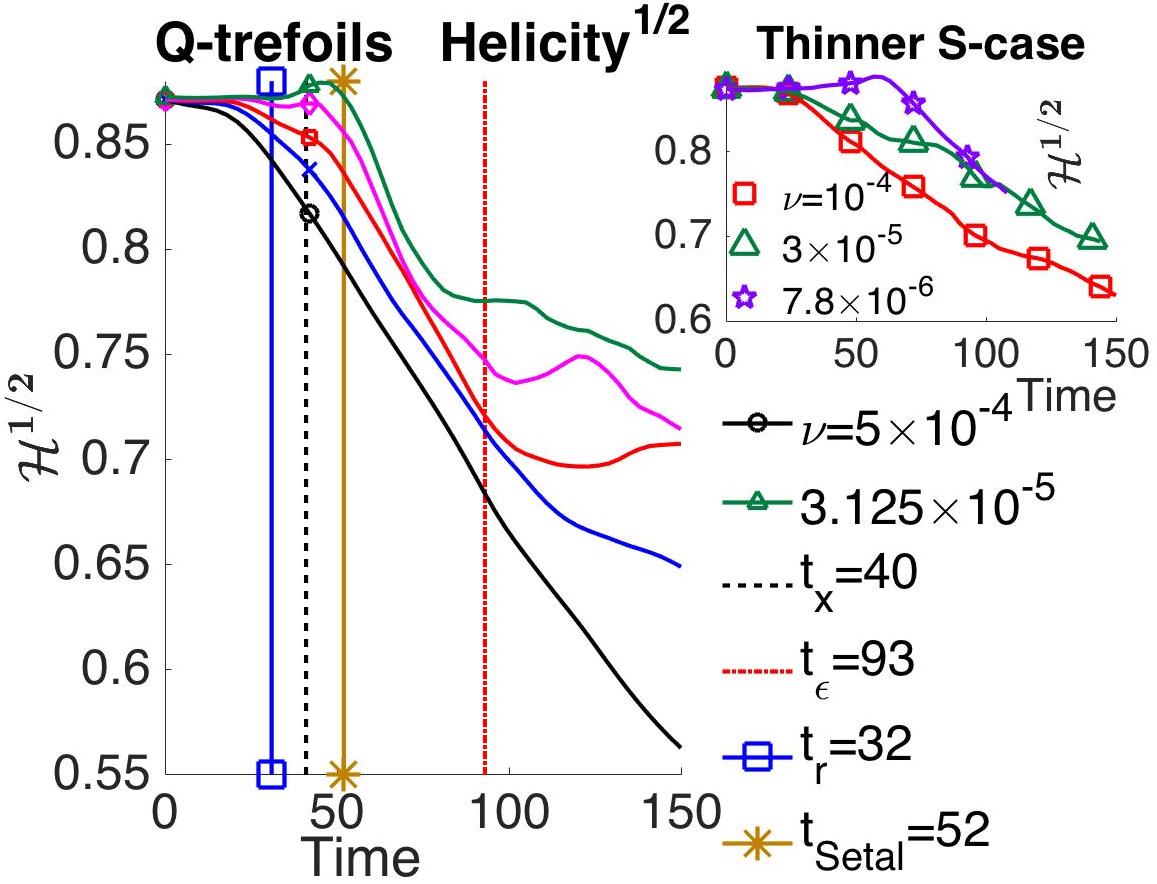}
\caption{\label{fig:HelQS} Time evolution of the scaled  helicity
${\cal H}^{1/2}$. In the main frame are 5 viscosities from $\nu=5\times10^{-4}$ to 
$3.125\times10^{-5}$ for the Q-cases.  Important times during the evolution of the 
trefoil: The blue-$\square$ line is $t=31$, the time when the first signs of reconnection 
are visible as in figure \ref{fig:T31}. The dashed line is $t_x=40$, 
the time when all the $\sqrt{\nu}Z$ meet and the first reconnection ends.
The brown-$\star$ line is $t_{Setal}=52$, roughly the equivalent time to when the 
\ScheeleretalIrvinefourt\yearpp{2014} experiment would end with respect to
choosing $t_x(SK)=638$ms. 
The red-dot-dash line is the Q-trefoil dissipation saturation time $t_\epsilon=93$ 
indicated in figure \ref{fig:QSsnuZ}. 
The upper-right inset shows ${\cal H}^{1/2}$ for three viscosities 
of the thinner core S-trefoils to show that the timescale for  helicity decay 
is independent of the core thicknesses $r_e$.  }
\end{figure} 

\section{Enstrophy evolution and timescales \label{sec:Z}}

Several nonlinear and viscous timescales can be applied to the vortex reconnection events. The
most important nonlinear timescale for the numerical trefoils is identified in 
figure \ref{fig:QSsnuZ} by the viscosity-independent crossing time of 
$B_\nu(t)=(\sqrt{\nu}Z(t))^{-1/2}$ \eqref{eq:Bnu} at $t=t_x\approx40$ for the
Q-trefoils and $t_x\approx45$ for the S-trefoils, indicating that $t_x$ is independent of $\nu$
and approximately independent of $r_e$. 
Figure \ref{fig:T42} at $t=42$, discussed in section \ref{sec:t40}, demonstrates that 
$t_x\approx 40$ represents when the first reconnection ends in physical space for the Q-trefoils. 

Nonlinear time and velocity scales that do not depend upon $\nu$ and $r_e$ can be formed using  
the circulation $\Gamma$ and the size of the structure $r_f$: 
\EQL{eq:ta} t_f=r_f^2/\Gamma\quad{\rm and}\quad v_f=\Gamma/r_f   \,.\EN 
For both the Q and S-trefoils, $t_f$(Q)=$t_f$(S)$=2^2/0.5=8$ and $v_f=0.25$. 
\RMK\yearpp{2017} concluded that $t_f=8$ and $t_x\approx40$ are related, despite 
$t_x\sim 5t_f$, by finding the self-similar collapse \eqref{eq:isnuZBxtime} 
that covers the linearly decreasing regimes of $B_\nu(t)$ in figure \ref{fig:QSsnuZ}.

For velocities, the nearly vertical maximum velocity is approximately $v_f$ and
up to $t=20$ the vertical velocity of $\bx_\infty$, the position of $\|\omega\|_\infty$, 
is $v_\infty\approx 0.25=v_f$.  This is discussed further in section \ref{sec:t40}.

\subsection{ Experimental timesscales \label{sec:exptime}}

How can equivalent reconnection times $t_x$ for the experiments of 
\KlecknerIrvinethirt\yearpp{2013} and \ScheeleretalIrvinefourt\yearp{}{2013} be determined?
Two routes for estimating the experimental reconnection timescales are considered. First,
the experimental nonlinear timescale $t_f$ \eqref{eq:ta} can be estimated if $\Gamma$, the 
circulation of the shed vortex, and $r_f$, the radius of the knot \eqref{eq:trefoil}, are known.  
Second, visual validation from the experimental reconnection figures, 
based upon when the first major reconnection has completed. 
There are problems with both approaches, so 
the $r_f=45$mm case from \KlecknerIrvinethirt\yearpp{2013} is discussed first
because the timescales of both methods can be estimated from the available data and figures. 

The problems are these. In the estimate of the experimental nonlinear timescale 
$t_f$ \eqref{eq:ta}, the published circulations $\Gamma$ were not measured, but were estimated 
based upon a flat-plate approximation of 
\EQL{eq:thinplate}\Gamma_{\mbox{fp}}=\pi \mbox{UC}\sin \alpha\,.\EN 
The basic parameters that determine $\Gamma_{\rm fp}$ are C, the chord (width) of the hydrofoil, U,
the velocity of the hydrofoil and the tilt $\alpha$ of the overall hydrofoil with respect to 
the direction of propagation. However, this neglects the contribution
due to for the highly curved ribbons used in their 3D-printed knot models. In aeronautics
this is known as the camber and is discussed in the next section.

The problem with defining $t_x$ as when the first reconnection ends using three-dimensional images is 
that this small, but significant, change in the topology cannot be identified from a single image 
in isolation, which can be interpreted in several ways. 
So multiple times need to be compared before a convincing, visually determined value for $t_x$ 
can be obtained. The strategy for addressing this problem will be to identify when a clear and 
persistent gap appears in the global trefoil structure. Then look backwards in time to 
the first earlier incident when along the trefoil vortex, there is a sudden local dispersion 
of the bubbles as lines meet and choose this as $t_x$. In figure \ref{fig:KI}, the clear gap 
is at $t=400$ms, and the earlier bubble dispersion event is at $t=350$ms.
This will be discussed further in section \ref{sec:t40}.

\subsection{Camber correction \label{sec:camber}}

To maximise the experimental circulation of their vortex knots, the 3D-printed knot models 
that \KlecknerIrvinethirt\yearpp{2013} and \ScheeleretalIrvinefourt\yearp{}{2014b} 
accelerated through their water tank used curved ribbons whose trailing edge was tilted 
$\theta=15^\circ=\pi/12$ rad. This corresponds to an angle of attack of $\alpha=\theta/2$.
The strong curvature or camber can be approximated as
\EQL{eq:camberh} y(x)=h\dfrac{x({\rm C}-x)}{{\rm C}^2/4}\quad{\rm with}\quad
\frac{h}{C}\approx\frac{\theta}{8}\,. \EN
If $x=({\rm C}/2)(1-\cos\phi)$ for $0\leq\phi\leq\pi$, the camber correction can then be 
approximated as $\Delta\Gamma=(\pi/2)\mbox{UC}\,{\rm A}_1$ where $A_1$ is determined by
the following integral of the derivative $dy/dx$ of the camber line 
\yearp{\HoughtonCarpenterthr}{2003} 
$${\rm A}_1=(2/\pi)\int_0^\pi(dy/dx)\cos\phi d\phi=
(2/\pi)(4h/{\rm C})\int_0^\pi\cos^2\phi d\phi=(4h/{\rm C})$$
giving 
\EQL{eq:DeltaGamma} \Delta\Gamma=(\pi/2)\mbox{UC}\,{\rm A}_1=2\pi(h/{\rm C}){\rm UC}
\approx\pi\mbox{UC}(\pi/48)\,. \EN

This represents a 50\% increase in the circulation on top of 
the primary flat plate contribution of $\Gamma_{\mbox{fp}}=\pi\mbox{UC}\pi/24$ for
$\alpha=\theta/2=\pi/24$ using \eqref{eq:thinplate}. Together 
\EQL{eq:flatcamber} \Gamma=\Gamma_{\mbox{fp}}+\Delta\Gamma=1.5\Gamma_{\mbox{fp}}=\pi^2\mbox{UC}/16 
\,.\EN

Once the camber-corrected $\Gamma$ are known, then the experimental nonlinear 
timescale $t_f$ \eqref{eq:ta} can be determined and its relationship with $t_x$ identified 
for comparisons with the simulations and the visible signs of experimental reconnection.  What 
are the $t_f$ for the \KlecknerIrvinethirt\yearpp{2013} and \ScheeleretalIrvinefourt\yearpp{2014}
experiments?

For the $r_f=45$mm experiment of \KlecknerIrvinethirt\yearpp{2013}, with $U=3.1$m/s and 
the chord C=15mm, the circulation is $\Gamma=2.8\times10^4$mm$^2$/s \eqref{eq:flatcamber}
and the timescale is $t_f$(KI)=$r_f^2/\Gamma=70$ms \eqref{eq:ta}. Based on the crossings of 
$B_\nu(t)$ at $t_x$ in figure \ref{fig:QSsnuZ}a, the end of the first reconnection should be at 
$t_x$(KI)$=5t_f$(KI)=350ms, which would be consistent with the conclusion of
\KlecknerIrvinethirt\yearpp{2013} and the discussion in section \ref{sec:t40} using 
the $t=350$ms and $t=400$ms frames in figure \ref{fig:KI} with {\it reconnection gaps}. 

For \ScheeleretalIrvinefourt\yearpp{2014}, 
\eqref{eq:flatcamber} with $r_f=69$mm, C=22.5mm and U=2m/s, gives $t_f$(SK)=168ms. However,
the predicted reconnection time of $=5t_f=840$ms is after the first clear gap
at $t\!\approx$638ms$\ll$840ms. This will be discussed further in section \ref{sec:Schetal}.

\begin{figure}
\includegraphics[scale=0.40]{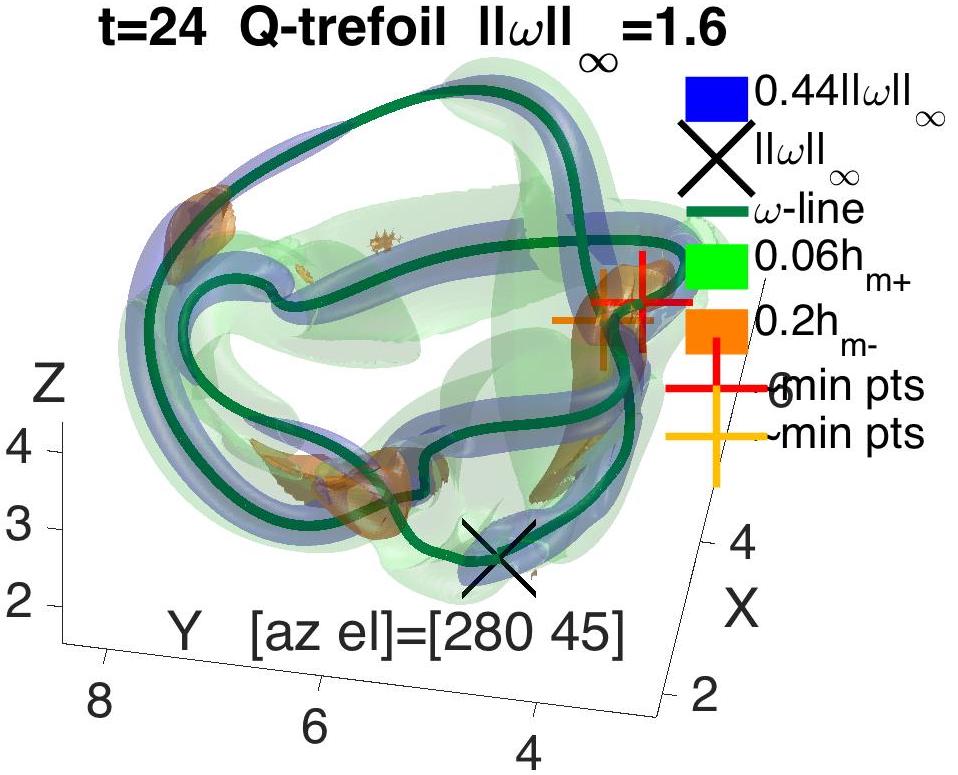}
\caption{\label{fig:T24} Isosurfaces at $t=24$, shortly before reconnection begins. The
vorticity isosurface is in blue, the position of $\|\bomega\|_\infty$ is {\bf X}, 
the trefoil line through it is green and the helicity isosurfaces are: $0.06\max(h)$ in green 
and $0.2\min(h)$ in orange where $\max(h)=-0.22$ and $\min(h)=0.085$.  
The points with the minumum distance between the two loops of the trefoil,
where reconnection is about to begin, are at the yellow and red {\bf +}'s. At the red {\bf +}, the
trefoil is bending back upon itself to become anti-parallel to the yellow {\bf +}. A kink
is forming on the trefoil line to the left of $\|\bomega\|_\infty$ in another region of $h<0$.
}
\end{figure}

\begin{figure} 
\includegraphics[scale=0.40]{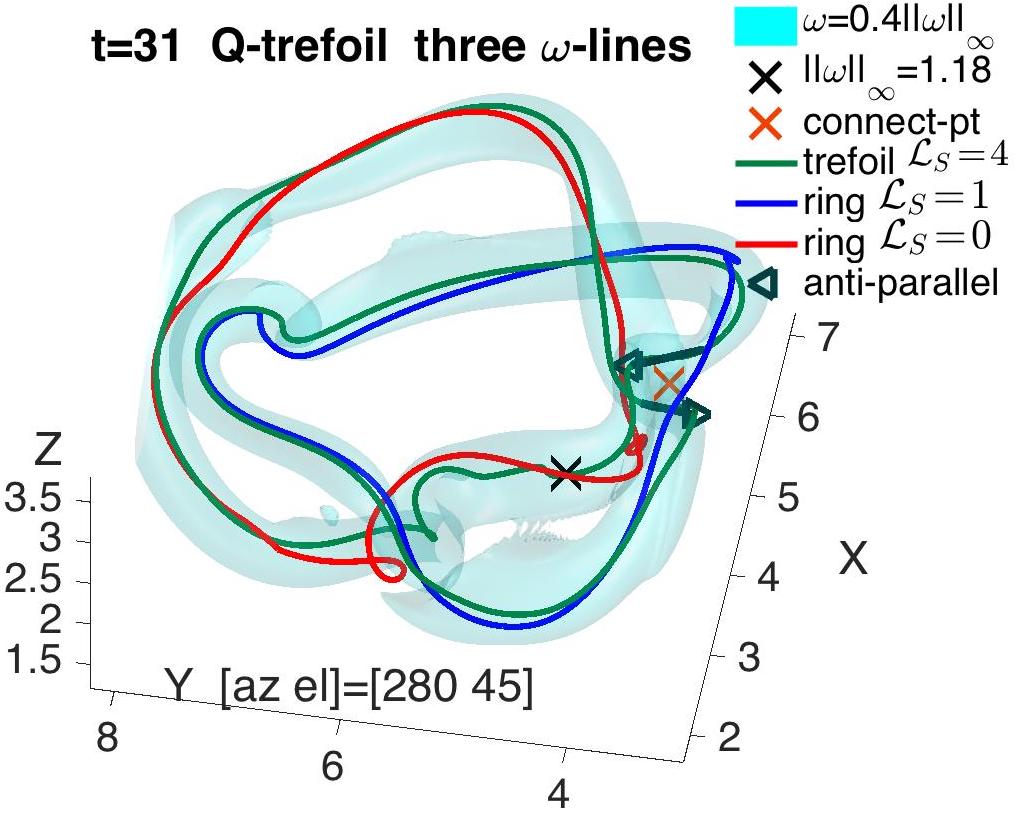}
\caption{\label{fig:T31} A single vorticity isosurface plus three closed vortex lines 
at $t=31$, the first time that visible reconnection is observed. 
The green trajectory follows a trefoil trajectory seeded near $\|\bomega\|_\infty$ 
indicated by {\bf X}.  The green trajectory's self-linking is ${\cal L}_S=4$, which can be 
split into $W\!r+T\!w=2.85+1.15=4$ At the closest approach of the trefoil's two loops, 
due to an extra twist, the loops are anti-parallel, as indicated by two arrows.
Between them is the {\it reconnection zone} whose mid-point is shown by the orange {\bf X}.  
Trajectories seeded on either side of this point, away from the trefoil, become linked rings.
For the red ring ${\cal L}_S=0$ and for the blue ${\cal L}_S=1$ with a 
total linking of ${\cal L}_{rb}+{\cal L}_{Sb}+{\cal L}_{Sr}=3$.}
\end{figure}

\begin{figure}
\includegraphics[scale=0.40]{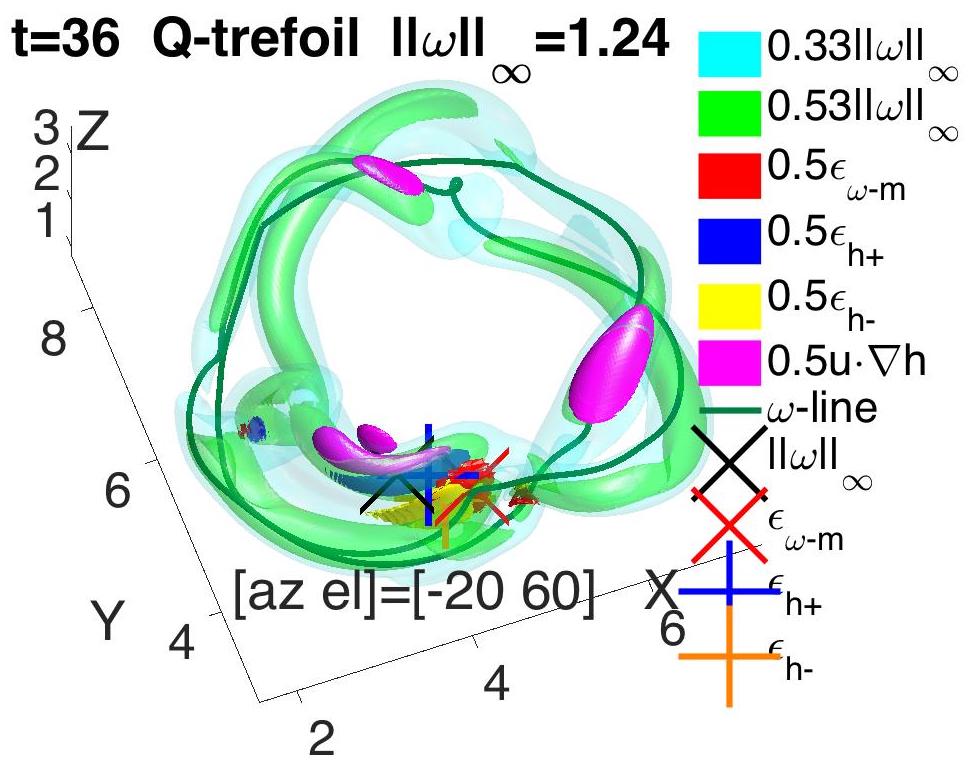}
\caption{\label{fig:T36} Multiple vorticity and dissipation isosurfaces at $t=36$ rotated 
90$^\circ$ clockwise from figure \ref{fig:T31} at $t=31$.  The black and red {\bf X}'s mark the 
positions of $\|\omega\|_\infty$ and maximum enstrophy dissipation 
$\epsilon_{\omega-{\rm m}}=\max(\epsilon_\omega)$ \eqref{eq:enstrophy}. The blue and yellow
plus signs show the extremes of the helicity dissipation \eqref{eq:helicity}, 
$\epsilon_{{\rm h+}}=\max(\epsilon_h)=4.9$ and 
$\epsilon_{{\rm h-}}=\min(\epsilon_h)=-3.2$ respectively.
The green vortex trajectory, a trefoil plus an extra loop, and
the lower $\omega=0.33\|\omega\|_\infty$ cyan vorticity isosurface 
mostly follow the original trefoil. A gap exists in the green $\omega=0.53\|\omega\|_\infty$ 
vorticity isosurface in the {\it reconnection zone} between two red enstrophy dissipation 
isosurfaces with $\epsilon_\omega=0.5\epsilon_{\omega-{\rm m}}$.
Around the black {\bf X} at $\|\omega\|_\infty$ and stacked with the left 
$\epsilon_\omega$ red isosurface, are flattened isosurface of 
positive helicity dissipation, negative helicity dissipation and the advection of
negative helicity, respectively as blue for $\epsilon_{h}=0.5\epsilon_{h+}$, 
yellow-orange for $\epsilon_{h}=0.5\epsilon_{h-}$ and magenta for transport of negative helicity
$\bu\cdot\nabla h=0.5\max(\bu\cdot\nabla h)$ that will be tied to the later  $t>t_x$, $h<0$ regions
discussed in section \ref{sec:latetimes}. 
}
\end{figure}

\section{Evolution of the topology of the initial reconnection \label{sec:reconnection}}

The purpose of this section is to outline and compare the structural changes during and after 
the first reconnection using three-dimensional numerical images from six times and 
experimental images at five times from \KlecknerIrvinethirt\yearpp{2013} in figure \ref{fig:KI}. 
The common graphical tools for all times will be at least one level of vorticity isosurfaces and one
vortex line.  Additional isosurfaces of the dissipation of enstrophy, both signs of
helicity, both signs of helicity dissipation and the helicity transport will be added to the
figures where appropriate. 

For completeness and later analysis, the discussion will begin with figure \ref{fig:T24} 
at $t=24$ to see what underlies the self-similar collapse that goes back to $t_\Gamma\approx15$, 
as implied by figure \ref{fig:QSsnuZ}, and to see the evolution of the twists that
set up the first visible reconnection at $t=31$ in figure \ref{fig:T31}. 

Figure \ref{fig:T36} at $t=36$ shows the dynamical terms during reconnection, whose locations
are compared with figures \ref{fig:T42} and \ref{fig:T45} at $t=42$ and 45 to show that
the $t=t_x\approx40$ crossing of $B_\nu(t)=(\sqrt{\nu}Z(t))^{-1/2}$ \eqref{eq:Bnu} in 
figure \ref{fig:QSsnuZ}a is 
when the first reconnection ends with the formation of complete gaps in the trefoil structure,
as \RMK\yearpp{2017} showed for anti-parallel reconnection. 

Figure \ref{fig:T63} at $t=63$ is included to complement the first clear signs of 
negative helicity density, $h<0$ in the outer regions of the $t=42$ and 45 figures and to
show how the structure as the helicity finally begins to decay.
The $t>6$ isosurfaces will now be discussed in temporal order.

\subsection{Evolution as reconnection starts \label{sec:3Dtime}}

{\bf Figure \ref{fig:T24} at $t=24$} shows a severely contorted structure shortly before the
time that visible reconnection begins.  The blue vorticity isosurface and green vortex line 
have the configuration of the trefoil with negative helicity, $h<0$, forming in three regions.  
One around where reconnection will begin, between the yellow and red {\bf +}'s at 
the points of closest approach of the two loops of the trefoil. Note how 
the loop bending under the red {\bf +} is becoming anti-parallel with the loop under 
the yellow {\bf +}. This $h<0$ region extends almost to the position of $\|\omega\|_\infty$, 
the black {\rm X}. Another $h<0$ region appears where the loops are crossing again
to the left of the black {\rm X}. And finally, an $h<0$ region is forming on the opposite side of the
trefoil from the black {\rm X} at $\|\omega\|_\infty$. This region will grow significantly at the 
late times in the figures for $t=42$, 45 and 63.

{\bf Figure \ref{fig:T31} at $t=31$} was chosen to show how, within the {\it reconnection zone}, 
the dissipation terms in \eqref{eq:enstrophy} generate infinitisimal partial reconnections that 
gradually convert the trefoil into linked trajectoris. The steps for identifying these loops begin
with finding the mid-point between the closest approach of the trefoil's two loops, identified by 
the orange {\bf X}. About this point, the trefoil loops are tangent and anti-parallel, 
as shown by the arrows on the loops. The self-linking number of the green trefoil curves, determined
by applying \eqref{eq:link} to two parallel trajectories, is ${\cal L}_{Sg}=4$ and 
is due to an extra acquired twist first noted in the $t=24$ figure.

Next, several vortex trajectories were seeded about the orange {\bf X}. The linked 
vortex loops shown in red and blue were seeded on oppsite sides of the orange {\bf X} in the
direction perpendicular to both the direction separating the loops of the trefoil and the tangents 
to the trefoil loops.

Since the red and blue loops are linked and the blue loop has twist+writhe whose self-linking 
is ${\cal L}_{Sb}=1$, the total linking number of the red and blue loops is 
${\cal L}_t={\cal L}_{rb}+{\cal L}_{br}+{\cal L}_{Sb}=3$ \eqref{eq:link}, equal to the 
total linking number of the original trefoil. This demonstrates why,
if helicity is simply ${\cal H}=\Gamma^2{\cal L}$ \eqref{eq:Hlink}, reconnection by itself 
need not result in a change in the total helicity \yearp{\Laingetalfivet}{2015}. 

{\bf Figure \ref{fig:T36} at $t=36$} shows the location and strength of the terms from the
enstrophy and helicity budget equations (\ref{eq:enstrophy},\ref{eq:helicity})
during the final stages of the first reconnection. 
The marked locations are the positions of $\|\omega\|_\infty$ (black {\bf X}), 
$\max(\epsilon_\omega)$ (red {\bf X}),
and two {\bf +} signs, blue and yellow, at the positions of the maximum and minimum 
of the helicity dissipation term. 
The primary green vorticity isosurface has several gaps, including 
one between two red dissipation of enstrophy $\epsilon_\omega=0.5\max(\epsilon_\omega)$ 
isosurfaces indicating the {\it reconnection zone}.  To complement the continuous vortex 
trajectory and show that most of the original trefoil profile still exists at this time,
there is an additional lower threshold vorticity isosurface in cyan.  

There are several isosurfaces showing the helicity dissipation term.
Positive helicity dissipation with $\epsilon_{h+}=0.5\max(\epsilon_h)=4.9$ is in blue and 
negative with $\epsilon_{h-}=0.5\min(\epsilon_h)=-3.2$ is in yellow/orange (B/W lightest grey-scale).
These regions balance one another, allowing helicity to be preserved during 
this reconnection in a manner consistent with the proposal of
\Laingetalfivet\yearpp{2015}. 
There are also several magenta surfaces for the transport of negative helicity that will be 
connected to the formation of $h<0$ regions in figure \ref{fig:T45} at $t=45$ and
discussed in section \ref{sec:latetimes}. 

Note the following for figure \ref{fig:T36}:
\ITM
\item The vortex trajectory was identified by applying \eqref{eq:vortexlines} to the point 
with the maximum of vorticity in the $x>\ell/2$ half domain and has an extra loop as it loses 
its way through the reconnection zone. 
\item The {\it reconnection zone}§ between the two red enstrophy dissipation \eqref{eq:enstrophy} 
isosurfaces with $\epsilon_\omega=0.5\max(\epsilon_\omega)$ is above the points $x=4$ and $x=5$ and
covers the advected location of the $t=31$ {\it reconnection zone} with locally anti-parallel
vorticity. 
\item Sandwiched together around $\|\omega\|_\infty$ (black {\bf X}), on the left 
side of the {\it reconnection zone}, are sheets of helicity and enstrophy dissipation and
$h<0$ helicity transport stacked in the following order from top-left to the bottom: 
Helicity advection, positive helicity dissipation, enstrophy dissipation and negative helicity 
dissipation. This suggests that there is an undetermined, underlying structure on that side of
the dissipation.  
\ITN

\subsection{Evidence that the first reconnection ends at $t=40$ \label{sec:t40}}

\begin{figure}
\hspace{10mm}
\includegraphics[scale=0.34]{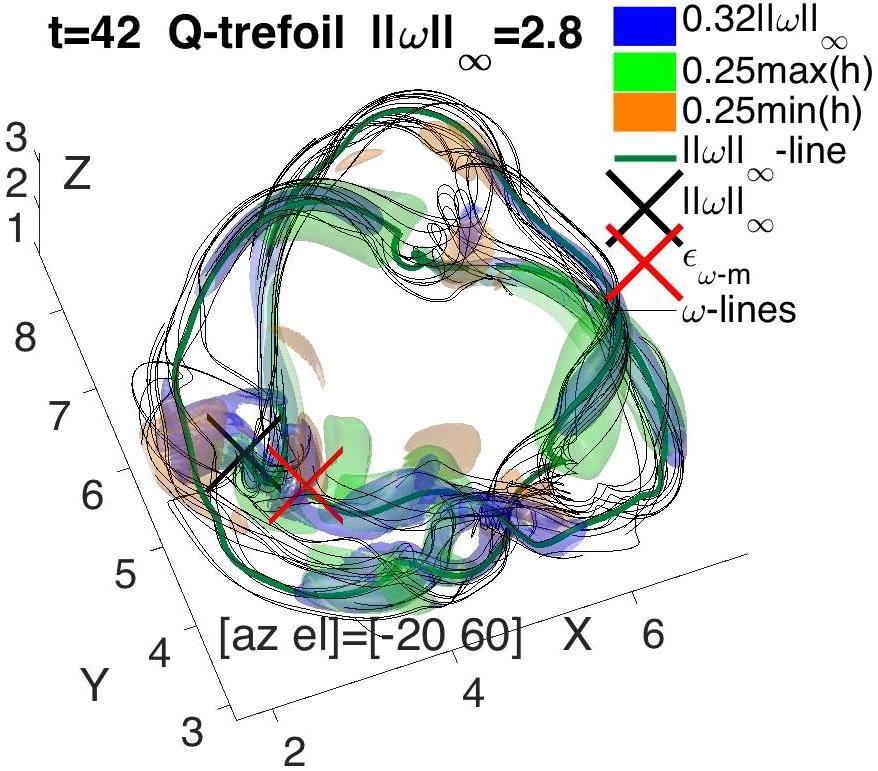}
\begin{picture}(0,0)\put(-160,139){{\LARGE\bf (a)}}\end{picture}
\\ \bminic{0.4}~\emini \bminic{0.6}\vspace{-14mm}
\includegraphics[scale=0.34]{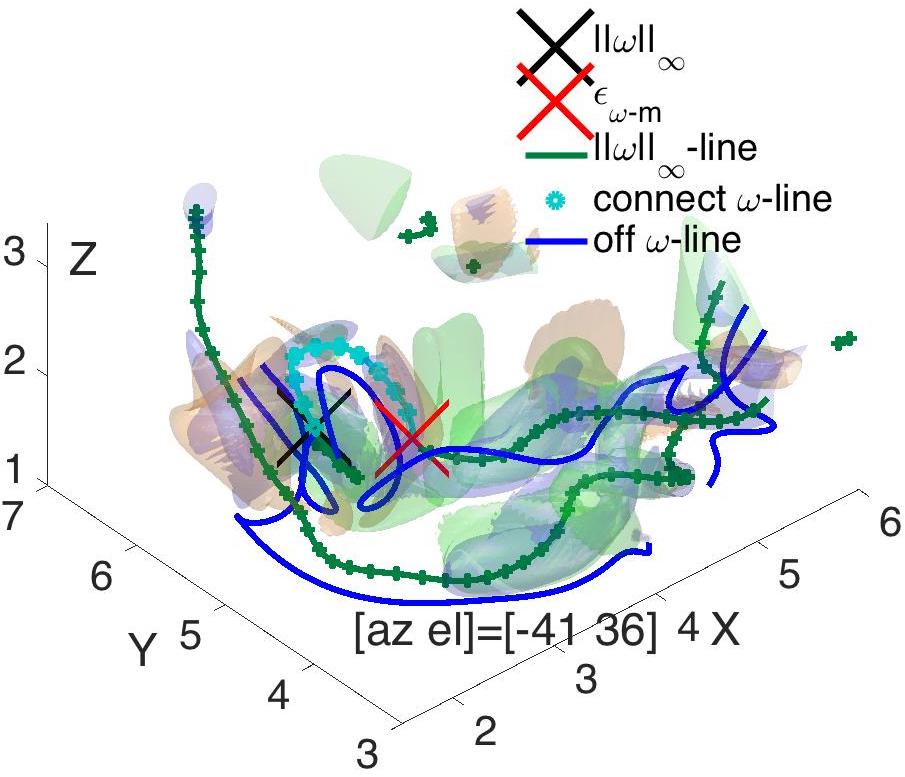}
\begin{picture}(0,0)\put(80,229){{\LARGE\bf (b)}}\end{picture}\emini
\caption{\label{fig:T42} Isosurfaces and vortex lines at $t=42$, just as the first
reconnection is ending.  Vorticity isosurfaces are in blue and the helicity isosurfaces are at 
$0.15\max(h)$ in green and $0.15\min(h)$ in yellow/orange, where $\max(h)=0.40$ and $\min(h)=-0.25$. 
{\bf a:} Besides a green trefoil line 
there are multiple thin, black vortex lines originating from the vicinity of 
the black {\bf X} at $\|\omega\|_\infty$. 
The region with the multiple dissipation surfaces in figure \ref{fig:T36} at $t=36$ is to
the right of the red {\bf X} at $\epsilon_{\omega-m}=\max|\epsilon_\omega|$ \eqref{eq:enstrophy}
indicating where strong enstrophy dissipation is continuing.
{\bf b:} Close-up of where there was reconnection. 
The two vortex lines are the green line through $\|\omega\|_\infty$ with 
cyan bullet marks that highlight its path between the black {\bf X} and 
the red {\bf X} and a vortex line in blue. The two lines, which
both originated near $\|\omega\|_\infty$, diverge from one another a bit before making detours 
around the gap in the largest blue vorticity isosurface close to the red {\bf X}. 
This gap represents the end of the first reconnection and then expands, as shown in 
figure \ref{fig:T45} at $t=45$.
} 
\end{figure}
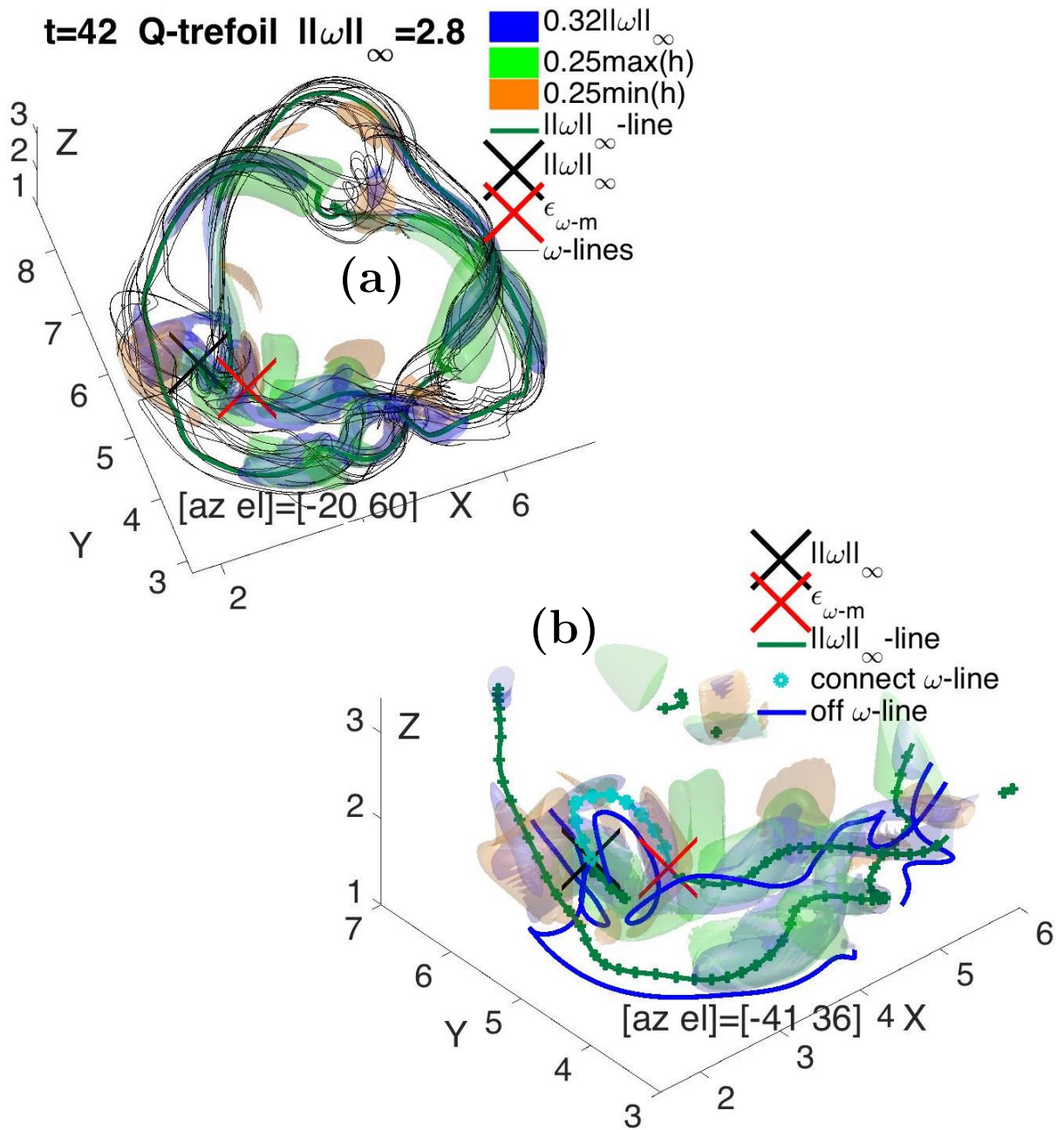

\begin{figure}
~~~~~~~~\bminic{0.45}\includegraphics[scale=0.75,clip=true,trim=0 110 0 10]{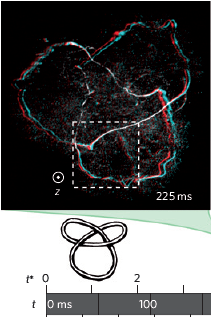}
\begin{picture}(0,0)\put(-0,128){$t=225$ms} \end{picture} \emini\bminic{0.1}~\emini\bminic{0.45}~
\includegraphics[scale=0.75,clip=true,trim=0 110 0 10]{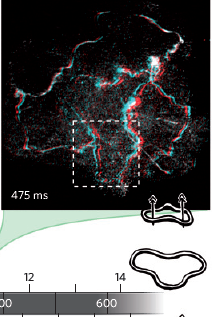}
\begin{picture}(0,0)\put(-0,128){$t=475$ms} \end{picture} \emini \\
\bminic{0.1}~\emini\bminic{0.8}
\includegraphics[scale=1.1,clip=true,trim=0 110 0 20]{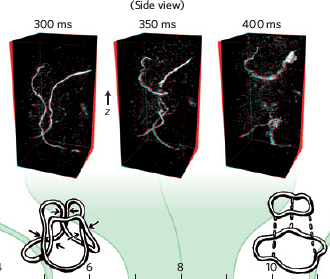}
\begin{picture}(0,0)\put(-23,114){\large Side} \put(-27,102){\large -views} 
\end{picture} 
\emini 
\caption{\label{fig:KI} Figures from \KlecknerIrvinethirt\yearpp{2013} at $t=225$ms, 300ms, 
350ms, 400ms and 475ms that correspond roughly to the $t=24$, 36, 42, 45 and 63 figures here. 
The $t=$300, 350, 400ms frames are side-view close-ups of the region that is outlined in the
$t=$225ms and 475ms frames.}
\end{figure}

\begin{figure}
\includegraphics[scale=0.36]{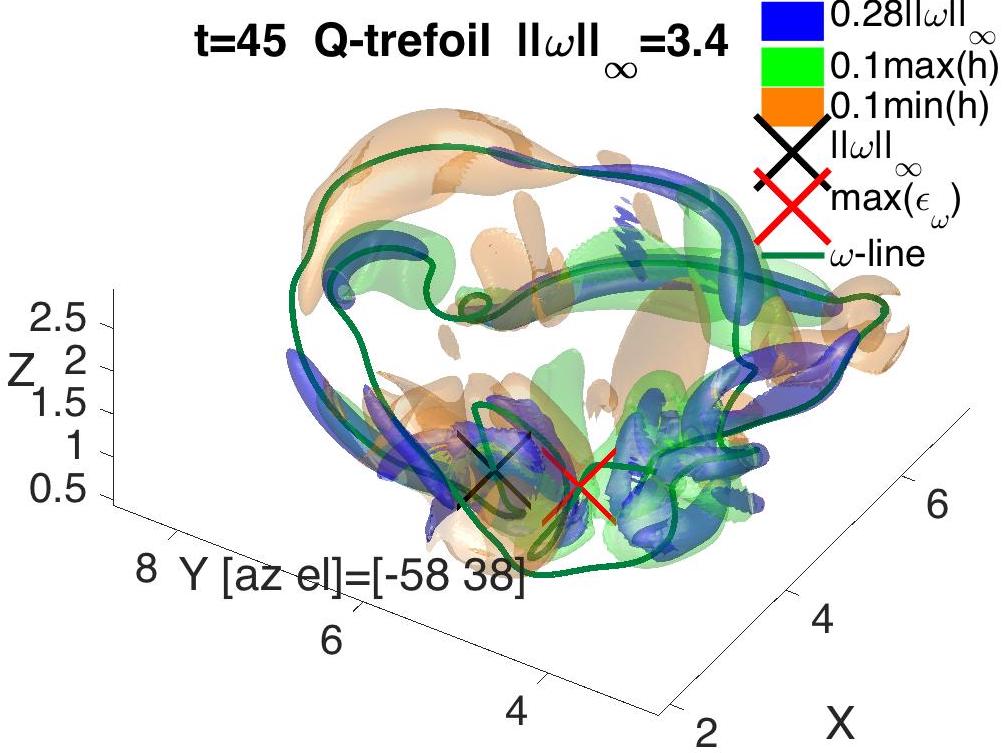}
\caption{\label{fig:T45} Isosurfaces and one vortex line at $t=45$ just after the first
reconnection has ended.  Vorticity isosurfaces are in blue and 
the helicity isosurfaces are at $0.1\max(h)$ in green and
$0.1\min(h)$ in yellow/orange, where $\max(h)=0.62$ and $\min(h)=-0.26$. 
A gap without strong vorticity, but twisted and bent vortices to either side, 
now covers the {\it reconnection zone} to the right of red {\bf X}.
Nonetheless, except in that zone, the vortex line seeded at the point of maximum vorticity 
at {\bf X} still has the flavour of the
original trefoil as it circumnavigates the centre twice and passes through regions 
with large vorticity and large helicity of both signs.  Green $h>0$ positive helicity overlying
twisted blue vorticity dominates to the right of the gap.  Yellow/orange $h<0$ negative helicity
dominates to the left of the gap between the black {\bf X} and red {\bf X}, with the strongest 
enstrophy dissipation \eqref{eq:enstrophy}, between blue isosurfaces of vorticity. 
There is another region of large negative helicity opposite to the reconnection zone in the 
upper left whose possible importance is discussed in section \ref{sec:latetimes}.
} 
\end{figure}

The visual evidence that the reconnection ends at $t=t_x=40$ will be based upon comparing
the structures at $t=36$, 42 and 45 in figures \ref{fig:T36}, \ref{fig:T42} and \ref{fig:T45}. 
Specifically, the changes between $t=36$ in figure \ref{fig:T36}, which shows us the dissipation terms
during reconnection, and the growing gap in the trefoil structure in figures 
\ref{fig:T42} and \ref{fig:T45} at $t=42$ and 45 respectively. Further evidence will be provided by
comparing the structures in figures \ref{fig:T31}, \ref{fig:T36}, \ref{fig:T42} 
and \ref{fig:T45} at $t=31$, 36, 42 and 45 with to the three frames taken from
\KlecknerIrvinethirt\yearpp{2013} in figure \ref{fig:KI} about their predicted reconnection time 
of $t_x$(KI)=350ms. 

{\bf Figure \ref{fig:T42} at $t=42$} has two frames, with the left frame showing with the entire 
trefoil and the right frame focussing upon the position where the stack of dissipation isosurfaces 
were found at $t=36$ in figure \ref{fig:T36}. This region is to the left of 
the {\it reconnection zone} from earlier times, which is now at the right end of this close-up. 
The two frames use the same isosurfaces, vorticity isosurfaces in blue, positive helicity 
isosurfaces at $0.15\max(h)$ in green and negative helicity isosurfaces at $0.15\min(h)$ 
in yellow/orange, where $\max(h)=0.40$ and $\min(h)=-0.25$.  Both frames also use the same
green vortex line through $\|\omega\|_\infty$. 

In the main frame, in addition to the green trefoil line through $\|\omega\|_\infty$, 
there are several thin, black vortex lines originating from the vicinity of $\|\omega\|_\infty$
that show where spirals are forming and help connect the remaining parts of 
the trefoil by creating bridges between the new negatively signed $h<0$ helicity regions 
and the original positively signed $h>0$ regions.  Spirals tend to form along these bridges.

The close-up uses only two vortex lines. The green $\|\omega\|_\infty$ line which is highlighted with
bullets of two colours, green and cyan, and an off-set line in blue. 
The green bullets show most of the 
$\|\omega\|_\infty$ line and the cyan bullets indicate its path between $\|\omega\|_\infty$ 
and $\epsilon_{\omega-m}$ (black and red {\bf X}'s).

Let us consider three features in the close-up. First, the dissipation of the enstrophy along
the lower branch of the two vortex lines where the dissipation isosurfaces were at $t=36$.
Second, the twisting of the vortex lines with green $h>0$ and blue enstrophy isosurfaces the right. 
This is the primary location for enstrophy growth. 

Third, where the cyan-bullets and the blue off-set line both avoid the space between the black
and red {\bf X's}. This is the first appearance of a complete gap in the reconnecing isosurfaces
and demonstrates that $t_x=40$ represents the end of the first reconnection, as suggested by
the $t\!=\!t_x\!=\!40$ crossing of $\sqrt{\nu}Z$ in figure \ref{fig:QSsnuZ}a. Another feature of this
location is that to the right, green $h>0$ dominates and the left, yellow/orange $h<0$ dominates,
which continues in figure \ref{fig:T45} at $t=45$.  Supporting this, if the close-up is 
rotated $90^\circ$ so that the right is on the bottom, the twisted ends resemble 
the $t=5t_f$(KI)$\approx$350ms frame in figure \ref{fig:KI} and the diagrams 
in figure 2 of \Laingetalfivet\yearpp{2015}. 

{\bf Figure \ref{fig:T45} at $t=45$}=5.6$t_f$(Q) shows the trefoil as the {\it reconnection gap} 
spreads between the position of $\epsilon_{\omega-m}\!=\!\max(\epsilon_\omega)$ (red {\bf X}) 
on the left and the blue, twisted vorticity isosurfaces to the right. 
Unlike earlier times, the green trefoil trajectory goes completely around this gap, similar to
how the vortex lines in the $t=400$ms$\approx5.6t_f$(KI) frame in figure \ref{fig:KI} avoid
the region where reconnection ended at $t=350$ms.  The separation between the green $h\!>\!0$
helicity isosurface on the left and the yellow/orange $h${$<$}0 helicity isosurface on the right
has also increased over the $t=42$ separation in figure \ref{fig:T42}b. 

Furthermore, the extent of all of the $\pm h$ isosurfaces grows, with the positive green helicity 
to the right of gap more obviously surrounding and covering 
the twisted blue vorticity isosurfaces where small-scale enstrophy is growing. 
To the left of the {\it reconnection gap}, near the {\bf X} at $\|\omega\|_\infty$, 
the negative yellow/orange helicity is between, but not on, 
the sharp bends in the blue vorticity isosurfaces and 
could be connecting to the $h<0$ yellow/orange helicity isosurfaces in the 
outer parts of the trefoil, as discussed in section \ref{sec:latetimes}.  

Now that the similarities at and post-reconnection between the simulation graphics at $t=42$ and
45 to the experiments at $t=350$ and 400ms have been identified, one can go back in time to look
for similarities between the $t$=300ms experimental image in figure \ref{fig:KI}
and the trefoil structures in the $t=31$ and 36 figures.
In particular, comparing where the loops clearly cross in the $t$=300ms image to the region
about the orange {\bf X} in figure \ref{fig:T31} at $t=31$, the location where its loops cross 
and bend back upon themselves in preparation for reconnection. 
Further back in time, where the experimental $t$=225ms 
frame has a kink on the left side of the highlighted box might be similar to the kink underneath 
the red and yellow {\bf +} signs in figure \ref{fig:T24} at $t=24$ which $h<0$ is flowing out of.

In summary, the steps in forming the {\it reconnection gap} are: 
\ITM\item At $t=36$ in figure \ref{fig:T36}, the gap begins to form
where the dissipation of both the vorticity and helicity, of both signs, is greatest.
\item After reconnection ends at $t\approx40$, figure \ref{fig:T42} at $t=42$ shows that 
there are twisted vortex lines on either side of the {\it reconnection gap} 
with only one vestigial piece of the original trefoil vortex that avoids the gap.  
\item And by $t=45$ in figure \ref{fig:T45}, the break in the original trefoil is largely complete.
\ITN
By all these measures, it is the formation of the {\it reconnection gap} that marks the end of 
the first reconnection and justifies using this timescale for comparisons with the experiments.
Furthermore, the formation of the {\it reconnection gap} marks
the beginning of the new phase of even stronger enstrophy growth that 
leads to the development of the {\it dissipation anomaly} starting at $t_\epsilon\approx 2t_x$ in
figure \ref{fig:QSsnuZ}a.

\subsection{\ScheeleretalIrvinefourt\yearpp{2014} timescales \label{sec:Schetal} }

As noted at the end of section \ref{sec:camber}, using $r_f$, C and U from 
\ScheeleretalIrvinefourt\yearpp{2014b}  in \eqref{eq:flatcamber} gives a nonlinear timescale
of $t_f$(SK)=168ms, which implies a reconnection time of $t_x=860$ms, 
after the experiment ends. However, using a visual time based upon comparing the frames from its 
S4 movie at $t=$596ms, 638ms and 658ms, times at which the bubbles marking the vortices disperse 
then reform, to the physical structures at $t=36$, 41 and 45 here, suggests that the end 
of the first reconnection should be at $t_x$(SK)=638ms. This is consistent with the
conclusion of \ScheeleretalIrvinefourt\yearpp{2014b} and is when they changed the colour of one
of the subsequent loops in the S4 movie.  
Based upon $t_x\approx 5 t_f$(SK) for the simulations, this would suggest that for the
\ScheeleretalIrvinefourt\yearpp{2014} experiment, the nonlinear timescale should be
$t_f({\rm SK})=128$ms. Inconsistent with the camber-corrected timescale. 

Is there a diagnostic that could be extracted from the available
experimental data that could resolve this inconsistency?  One measurement would be the initial
vertical motion of the entire trefoil structure, whose value for the Q-trefoils agrees with the 
estimate given by \eqref{eq:ta}. This could be extracted from the experimental trefoil movies.
Another possibility is that there is a dynamically viscous timescale at $t_x$ 
that depends upon the thicknesses $r_e$ of the filaments. For example
$\delta t_\nu=T_c(\nu)-t_x$ using the linearly extrapolated $T_c(\nu)$ from figure \ref{fig:QSsnuZ},

\begin{figure}
\includegraphics[scale=0.40]{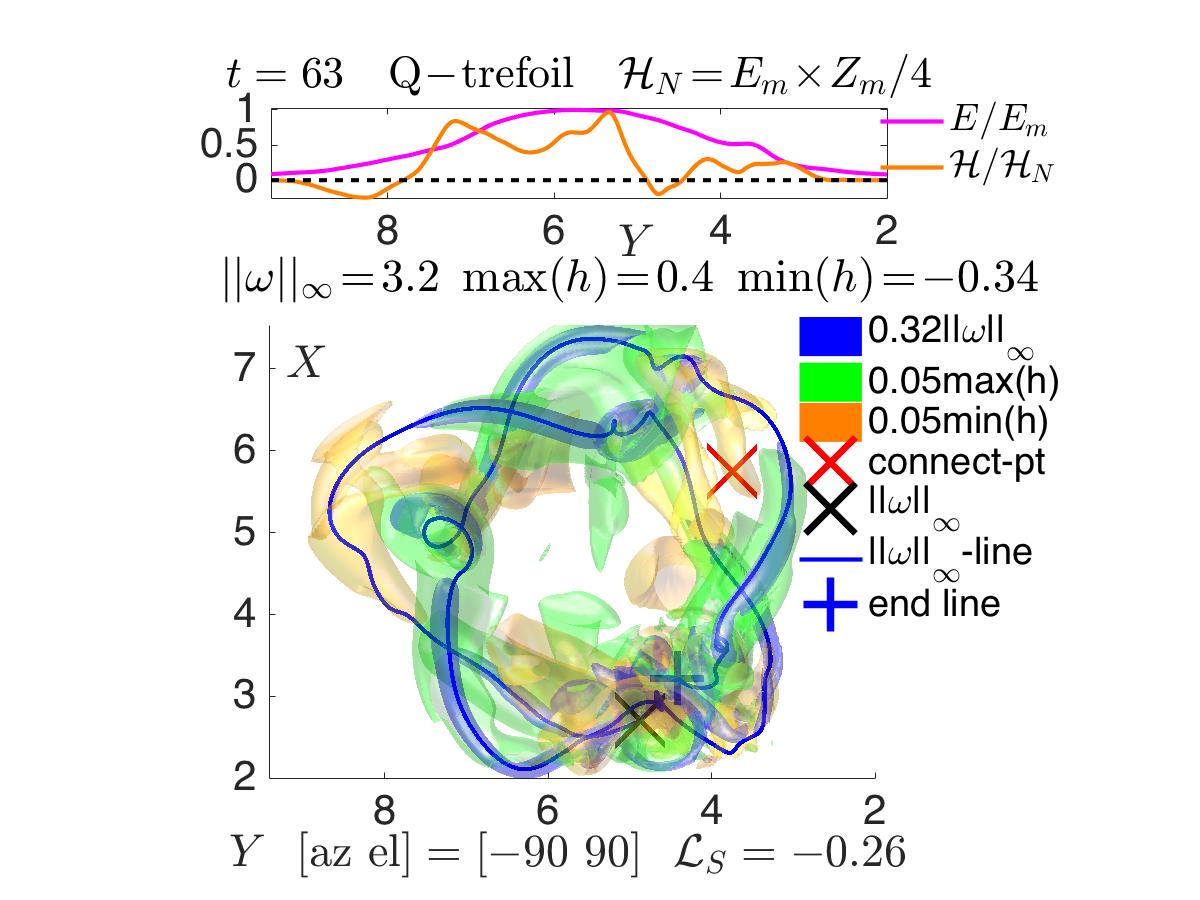}
\caption{\label{fig:T63} Profiles in $y$ of ${\cal H}(y)$ and $E(y)$ \eqref{eq:HEy} 
and isosurfaces at $t=63$ as reconnection is ending. $E_m=\max(E(y))$ and $Z_m=\max(Z(y))$
\eqref{eq:HEy}.
The isosurfaces are for vorticity in blue, with $0.32\|\omega\|_\infty$ 
where $\|\omega\|_\infty=3.2$ is at the black {\bf X}.  $h>0$ isosurfaces in green are for 
$0.05\max(h)$ where $\max(h)=-0.41$.
$h<0$ isosurfaces in yellow/orange are for $0.05\min(h)$ where $\min(h)=-0.34$.  
The advected position where reconnection began is indicted by the red {\bf X} in a gap
between blue $|\omega|$ and green $h>0$ isosurfaces. A blue vortex line that was seeded at
$\|\omega\|_\infty$ is followed for two circumnavigations of the trefoil before it was
terminated at the blue {\bf +} near $\|\omega\|_\infty$. Using \eqref{eq:link} with two
parallel trajectories, its non-integer self-linking number is ${\cal L}_S=-0.26$. 
}
\end{figure}

\subsection{Late times and negative helicity \label{sec:latetimes}}

The $t\leq t_x$ evolution not only generates the first reconnection, it also leads to the
next stage during which the global helicity ${\cal H}$ begins to decay slowly and the dissipation 
$\epsilon=\nu Z$ saturates, possibly generating a finite time, viscosity-independent
{\it dissipation anomaly}.
The following discussion using figures \ref{fig:T45} and \ref{fig:T63} at $t=45$ and 63 
is a first step in determining the physical structures during this period. The emphasis will be
on localised negative helicity $h<0$, its growth and role, more than the global ${\cal H}$ and
$\epsilon$.

First, a review of $h<0$ creation for $t\leq 45$.
Production of $h<0$ was noted as a precursor to reconnection at early times using 
figure \ref{fig:T24} at $t=24$ and then at $t=36$ in figure \ref{fig:T36} it was shown that the
dissipative production of $h<0$ compensated for the dissipative growth of $h>0$. Isosurfaces of 
the transport of $h<0$, $\bu\cdot\nabla h$ leading out of the reconnection zone were also noted. 
In section \ref{sec:t40} it was noted how yellow/orange $h<0$ forms to the left of the 
reconnection gap and $h>0$ dominates the region to its right in the $t=42$ and 45 figures.

Those observations can be connected to the continuing growth of the enstrophy $Z$ for 
$t\!>\!t_x\!=\!40$ as follows.  To the right at $t=45$, the twisted blue isosurfaces of enstrophy 
that are within a growing green envelope of $h>0$ are reminsicent of the post-reconnection swirling 
vortices seen for anti-parallel reconnection \yearp{\RMK}{2013}. 
However, since the global helicity ${\cal H}$ is preserved and therefore could suppress
the growth of $h>0$ associated with the enstrophy growth, the formation of the following regions
of $h\!<\!0$ relax this constraint. First, to the left of $\|\omega\|_\infty$ where the effect of 
the $h\!<\!0$ dissipation production region shown in figure \ref{fig:T36} at $t=36$ would be most 
immediate, $h<0$ isosurfaces form between bends in the enstrophy surfaces. Due to those bends,
$h<0$ is not tied to the local enstrophy $Z$ and therefore cannot accumulate, 
so transport of $h<0$ out of the vicinity of the {\it reconnection gap} is needed.

Second, transport of $h<0$ can be identified by carefully comparing 
where in figure \ref{fig:T36} at $t=36$ there are $\bu\cdot\nabla h>0$ isosurfaces for the
transport of $h<0$ and where in the $t=42$ and 45 figures there are $h<0$ regions 
outside the {\it reconnection zone}.  This includes a small 
$\bu\cdot\nabla h>0$ isosurface at the top of figure \ref{fig:T36}, which seems too
insignificant to account for the large $h<0$ yellow/orange region in the equivalent location
at $t=45$ in the upper left of figure \ref{fig:T45}.  The origin of this negative helicity 
could be the vorticity transport term in \eqref{eq:helicity} a term that originates from the
vortex stretching term in \eqref{eq:omega}. 
In figure \ref{fig:T42}a at $t=42$, vortex stretching 
is indicated by how the outer loop at the top has separated from the rest of the trefoil.

{\bf Figure \ref{fig:T63}} at $t=63$ uses $y$-profiles of the helicity, energy and enstrophy,
(\ref{eq:energy}, \ref{eq:enstrophy}, \ref{eq:helicity}) 
\EQL{eq:HEy} {\cal H}(y)=\int h dx\,dz,\quad E(y)=\int e dx\,dz, \quad{\rm and}\quad 
Z(y)=\int |\omega^2|  dx\,dz \EN
in the upper frame and a top-down three-dimensional perspective in the main frame to show the 
trefoil as the helicity is beginning to decay and the enstrophy growth is saturating. 
The upper frame shows that $E(y)$ and ${\cal H}(y)>0$ are concentrated in the centre 
of the trefoil. There are two regions of ${\cal H}(y)<0$, one off-centre to the right and 
one to the left at $y\approx8$.

Note the following two properties of the lower three-dimensional image.  Where reconnection began 
in figure \ref{fig:T31} at $t=31$ (red {\bf X}), there is at $t=63$ a large gap in the enstrophy 
and $h>0$ isosurfaces whose disappearance could represent the beginning of the decay of the large
positive global helicity ${\cal H}$ and saturation of the growth the enstrophy $Z$.  
The other property
is the significant $h<0$ isosurface along the outer loop to the left. This is the continuation of 
large, outer $h<0$ region noted at $t=42$ and 45, which combined with the $y$-profile at the 
top clearly shows that this $h<0$ region is outside the original envelope of the trefoil.  

This $y>8$, $h<0$ region is tenuously connected to what was the {\it reconnection zone} 
(red {\bf X}) by the blue vortex line that originates at the new position of 
$\|\omega\|_\infty$ at the bottom the figure, runs through the yellow $y>8$, $h<0$ region then
through the vicinity of the red {\bf X} twice.  This supports the evidence from figure \ref{fig:T42}a 
that the origin of this negative helicity region could be the vorticity transport term 
in \eqref{eq:helicity}. 

Although this single vortex line at $t=63$ retains the basic features of the trefoil, $t=63$ 
represents one of the last times that the overall trefoil structure can be seen as it breaks apart.
Overall, this is similar to the $t=$475ms frame of figure \ref{fig:KI}.  Note that this line 
terminates at the blue ({\bf +}) that is near, but not at, $\|\omega\|_\infty$ ({\bf X}), 
and has many twists. The result is that its self-linking number is ${\cal L}_S=-0.26$, a small, 
non-integer. 

\section{Conclusion \label{sec:conclude}}

This paper has presented the structural transformation of a trefoil vortex knot from the first signs 
of reconnection, past when that reconnection finishes and on to when helicity decay
and finite dissipation begins.

Previously, the experiments indicated preservation of the centreline helicity despite a change in 
the topology. The simulations in \RMK\yearpp{2017} supported this conclusion by tracking the global 
helicity ${\cal H}$ \eqref{eq:helicity}
as the trefoil loops began to reconnect. In addition, a new enstrophy scaling 
regime using $B_\nu(t)=(\sqrt{\nu}Z(t))^{-1/2}$ \eqref{eq:Bnu} was identified over this period
with $\nu$-independent crossing of $B_\nu(t)$ at a fixed time $t_x$. 
Self-similar $\nu$-independent collapse using \eqref{eq:isnuZBxtime} was then found
for both the Q-trefoils and new anti-parallel reconnection calculations \yearp{\RMK}{2017}. 
The anti-parallel calculations showed that $t_x$ is also when the first reconnection ended 
and for the Q-trefoil, figures \ref{fig:T36}, \ref{fig:T42} and \ref{fig:T45} at 
$t=36$, 42 and 45 here show that $t_x$(Q)=40 is
also when the first reconnection of the Q-trefoils ends.

Besides clearly identifying the structure of the trefoil as reconnection ends, 
the goal here has been to examine the dynamics underlying the evolution of these global properties.
Evidence for how viscous reconnection eats through the original trefoil loops is demonstrated
using an example of how newly reconnected vorticity can be generated
in figure \ref{fig:T31} at $t=31$
and by the close-up in figure \ref{fig:T42}b with vortex lines avoiding
a newly created gap in the vorticity isosurfaces. This leads to a clear {\it reconnection gap} 
forming at $t=45$ in figure \ref{fig:T45}, similar to the gaps that are used to identify equivalent 
reconnection times in the two experiments considered.

The physical space relationships between the diagnostic terms in the enstrophy and helicity budget
equations (\ref{eq:enstrophy},\ref{eq:helicity}) are used to tie together different aspects of this 
single phenomena and to investigate what role the unexpected preservation of helicity has in 
generating the new enstrophy scaling regime. From this,
it is found that negative helicity plays a role in every step. This starts during the
re-alignment of the trefoil loops before physical reconnection begins in figure \ref{fig:T24}
at $t=24$ and continues with the identification of 
regions of oppositely signed helicity dissipation in figure \ref{fig:T36} at $t=36$ that can
explain why the global helicity ${\cal H}$ can be preserved despite a change in the topology.

During this process, $h<0$ is created not only by the dissipative terms, but also by the
vorticity transport term and advection in \eqref{eq:helicity}.  The 
generation of large-scale negative helicity is demonstrated at $t=45$ and 63 in figures
\ref{fig:T45} and \ref{fig:T63} and appears to be a necessary condition for the 
enstrophy to continue to grow within the original envelope of the trefoil as simulations 
with decreasing $\nu$ are run. One can view this as an exhaust mechanism that allows the 
small-scale vorticity, enstrophy and positive $h>0$ helicity to cascade together
to ever smaller scales without being suppressed by the global helicity ${\cal H}$. 
A dynamical property that would not have been noticed without the experimental results 
on helicity preservation \yearp{\ScheeleretalIrvinefourt}{2014}.

This process continues until the trefoil structure finally begins to break apart at $t=$63 
in figure \ref{fig:T63}, which is roughly when the global helicity ${\cal H}$ begins to decay
in figure \ref{fig:HelQS} and the dissipation rate $\epsilon=\nu Z$ saturates 
in figure \ref{fig:QSsnuZ}.

Does this occur in the experiments? 
To cement the connection with the experiments, a strong correspondence between the evolution of 
the simulated Q-trefoil and the graphics for the earlier \KlecknerIrvinethirt\yearpp{2013} 
experiment in figure \ref{fig:KI} has been demonstrated.

Finally, if the $h<0$ exhaust is ever impeded by the periodic boundaries, then we have a
physical mechanism for suppressing for enstrophy growth for flows with strong periodicity or
symmetries that complements the mathematical bounds proven by \Constantineightsix\yearpp{1986}. 
Fortunately, those bounds can be relaxed simply by increasing the size $\ell$ of 
the periodic $\ell^3$ domains to far beyond the traditional $(2\pi)^3$ domain \yearp{\RMK}{2017}
and it seems plausible that unbounded growth of the enstrophy $Z$ is allowed as $\nu\!\to\!0$,
which could allow the formation of a {\it dissipation anomaly}. That is finite energy 
dissipation in a finite time as $\nu\to0$ from smooth solutions, without invoking singularities 
or roughness.

\section*{Acknowledgements}

I wish to thank S. Schleimer at the University of Warwick and H. K. Moffatt at Cambridge University
for clarifying the meaning of writhe, twist and self-linking. 
This work has also benefitted from conversations at the 2016 IUTAM events in Venice and Montreal. 
Computing resources have been provided
by the Centre for Scientific Computing at the University of Warwick, including use of the
EPSRC funded Mid-Plus Consortium cluster.

\section*{References}
\begin{harvard}





\item[] \authtwo{L.}{Biferale}{R.M.}{Kerr}
\yjour{1995}{Phys. Rev. E}{52}{6113}{--6122}
{On the role of inviscid invariants in shell models of turbulence}




\item[] 
\authone{G.}{Calugareanu}
\yjour{1959}{Res. Math. Pures Appl.}{4}{5}{--20}{L'int\'egral de Gauss et l'analyse
des noeuds tridimensionels}

%

\item[] 
\authone{P.}{Constantin}
\yjour{1986}{Commun. Math. Phys.}{104}{311}{--326}
{Note on Loss of Regularity for Solutions of the 3—D Incompressible Euler
and Related Equations}

\item[] 
\authtwo{E.L.}{Houghton}{P.W.}{Carpenter}
\ybook{2003}{Aerodynamics for Engineering Students, 5th ed.}{Butterworth-Heinemann}.







\item[] 
\authone{R.M.}{Kerr}\yjour{2005}{Fluid Dyn Res}{36}{249}{--260}
{Vortex collapse and turbulence}



\item[] 
\authone{R.M.}{Kerr}\yjour{2013}{Phys. Fluids}{25}{065101}{}{Swirling, 
turbulent vortex rings formed from a chain reaction of reconnection events}




\item[] 
\authone{R.M.}{Kerr}\yjour{2017}{J. Fluid Mech.}{submitted}{}{}
{Scaling of a perturbed Navier-Stokes trefoil}

\item \authfour{C.}{Rorai}{J.}{Skipper}{R.M.}{Kerr}{K.R.}{Sreenivasan}
\yjour{2016}{J. Fluid Mech.}{808}{641}{--667}
{Approach and separation of quantised vortices with balanced cores}






\item[] \authtwo{D.}{Kleckner} {W.T.M}{Irvine}
\yjour{2013}{Nature Phys.}{9}{253}{--258}{Creation and dynamics of knotted vortices}

\item[] 
\auththr{C. E.}{Laing}{R.L}{Ricca}{D.W.L.}{Sumners}
\yjour{2015}{Sci. Rep.}{5}{9224}{}
{Conservation of writhe helicity under anti-parallel reconnection.}




\item[] 
\authone{H.K.}{Moffatt}\yjour{1969}{J. Fluid Mech.}
{35}{117}{--129}{Degree of knottedness of tangled vortex lines}

\item[] 
\authone{H.K.}{Moffatt}\yjour{2014}{Proc. Nat. Acad. Sci.}{111}{3663}{--3670}{Helicity and singular structures in fluid dynamics}

\item[] 
\authtwo{H.K.}{Moffatt}{R.}{Ricca}
\yjour{1992}{Proc. Roy. Soc. Math. Phys. Eng. Sci.}{439(1906)}{411}{-–429}
{Helicity and the Calugareanu invariant}

\item[] 
\auththr{G.}{Sahoo}{F.}{Bonaccorso}{L.}{Biferale}
\yjour{2015}{Phys. Rev. E}{92}{051002}{}
{Role of helicity for large- and small-scales turbulent fluctuations}


\item[] \authmanytwo{M. W.}{Scheeler}{D.}{Kleckner}
\auththr{D.}{Proment}{G. L.}{Kindlmann}{W.T.M.}{Irvine}
\yjour{2014}{Proc. Nat. Acac. Sci.}{111}{15350}{--15355}{Helicity conservation 
by flow across scales in reconnecting vortex links and knots.}

\item[] 
\authmanytwo{M. W.}{Scheeler}{D.}{Kleckner}
\auththr{D.}{Proment}{G. L.}{Kindlmann}{W.T.M.}{Irvine}
Supporting information for \ScheeleretalIrvinefourt\yearpp{2014}. www.pnas.org/cgi/content/short/1407232111.




\end{harvard}

\end{document}